THE UNIVERSITY OF NOTTINGHAM

MALAYSIA CAMPUS

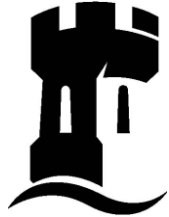

# NUMERICAL MODELLING OF SURFACE WAVE PACKET EVOLUTION ON VARYING TOPOGRAPHY

A dissertation submitted for the fulfilment of the requirement for the degree

Master of Philosophy

in

Engineering

by

Jieqiang Tan

# Abstract

An initial value problem of the one-dimensional nonlinear Schrödinger (NLS) equation with constant dispersive and nonlinear coefficients can be solved using a compact finite difference scheme (Xie, Li, & Yi, 2009). A similar scheme is implemented in the signalling problem of the one-dimensional NLS equation with constant coefficient where it describes the propagation of surface gravity wave packet over flat bottom. Various examples are illustrated.

A similar compact finite difference scheme is modified and implemented further in the signalling problem of the one-dimensional NLS equation with variable dispersive and nonlinear coefficients which models the surface wave packet propagation over slowly varying bottom. Several topographies are considered and illustrated. It is observed that the wave envelope signal propagates opposed to topography height. The possibility of shaping topography near shores area to minimise the effect of high amplitude solitary waves are also examined.

# Declaration

The work described herein was undertaken in the Department of Applied Mathematics, Faculty of Engineering, the University of Nottingham Malaysia Campus between September 2009 and August 2010. This thesis is my own original work except where specifically indicated.

# Acknowledgement

This research work is carried out under supervision of Dr. Natanael Karjanto from Department of Applied Mathematics and Prof. Andy Chan from Department of Civil Engineering. The New Researchers Fund NRF 5035, from the University Park Campus - University of Nottingham and the University of Nottingham Malaysia Campus is gratefully acknowledged.

# Table of Contents

# List of Figures

# Symbols and Notations

## Notation   Description

*f(t)*         Arbitrary function of integration

**F**          Body force vector per unit mass

*a*            Constant dispersive coefficient

*b*            Constant nonlinear coefficient

$\rho$         Density of the fluid

$\delta$       Dimensionless variable for mean depth to horizontal length

$\varepsilon$  Dimensionless variable for surface wave amplitude to mean depth

*j*            Discretised spatial node

*h*            Discretised spatial step

*n*            Discretised time node

$\tau$         Discretised time step for the NLS equation with constant coefficients

$\phi$         Discretised time step for the NLS equation with variable coefficients

$\mu$          Dissipative coefficient

$\eta$         Free surface elevation function

*g*            Gravitational acceleration

$c_g$          Group velocity

$\upsilon$     Kinematic viscosity

$\psi$         Modulated wave packet

*H*            Nondimensional depth

$\Omega$       Potential Function

*P*            Pressure at any point in the fluid

*R*            Radii of curvature of the free surface

$\kappa$       Real field (mean flow)

| | |
|---|---|
| $T$ | Surface tension |
| $t$ | Time |
| $\alpha$ | Variable dispersive coefficient |
| $\beta$ | Variable nonlinear coefficient |
| $u$ | Velocity component x |
| $v$ | Velocity component y |
| $w$ | Velocity component z |
| $\varphi$ | Velocity Potential |
| $\mathbf{u}$ | Velocity vector of the fluid |
| $\boldsymbol{\omega}$ | Vorticity vector |
| $\lambda$ | Wavelength |
| $k$ | Wavenumber |

# 1 Introduction

## 1.1 Introduction to water waves

In this chapter, we present a short introduction to the theory of water waves, from the basic physical appearance of water waves to the mathematical model of propagation of water waves. The development of water waves theory and research will be presented alongside with the relation to the governing equations in fluid mechanics.

The theory of water waves has been a source of interesting mathematical problems after the early work by Sir Isaac Newton (Newton, 1687). Then the linear theory of water waves was advanced by Pierre-Simon, marquis de Laplace (Laplace, 1776), Joseph-Louis Lagrange (Lagrange, 1781), Siméon-Denis Poisson (Poisson, 1816) and Augustin-Louis Cauchy (Cauchy, 1827). Then the nonlinear wave theory was considered by Jozef Gerstner (Gerstner, 1802). The subsequent contributions by Sir George Gabriel Stokes (Stokes, 1846) and others in water waves were made in the nineteenth century (Craik, 2004).

The theory of water waves involves the equations of fluid mechanics, the concept of wave propagation and the important role of boundary conditions. It is perhaps, one of the most difficult areas in applied mathematics due the nonlinear complexity of the mathematical model of equations. In fact, most of the fundamental concept for nonlinear dispersive waves and solitons are originated from the investigation of water waves.

Back to the most basic concept, what are water waves? Water waves are generated due to disturbances spreading through the water surface. One common example widely used to illustrate wave phenomena is by throwing a stone into a still pond (see Figure 1.1). The stone creates a disturbance on the pond surface.

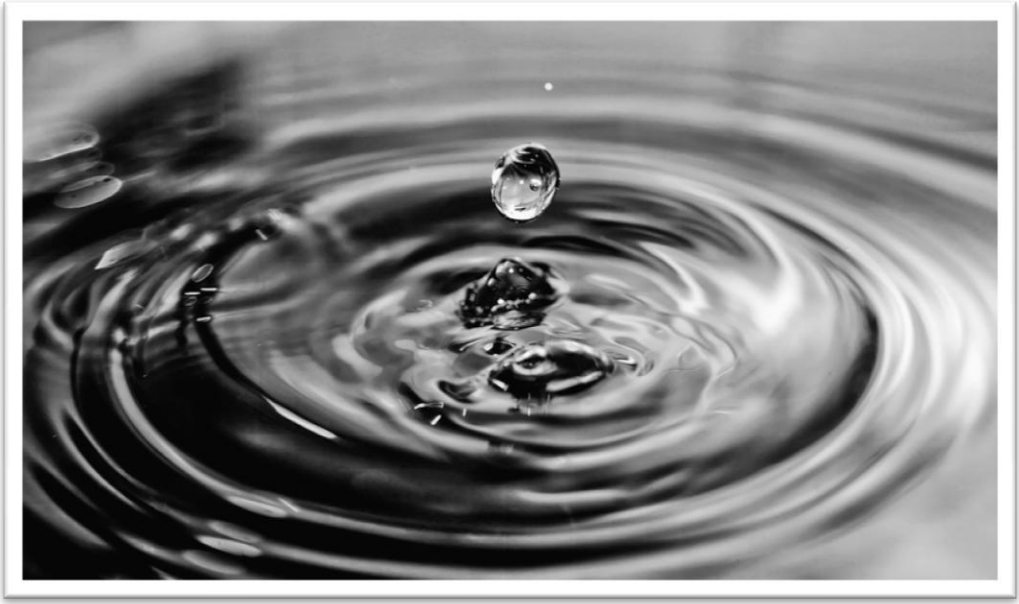

**Figure 1.1: Disturbance of a still water surface.**

Crests and troughs are formed as the disturbance is transferred horizontally through the surface of water. In other words, water waves are oscillations of water molecules propagate in horizontal direction (see Figure 1.2).

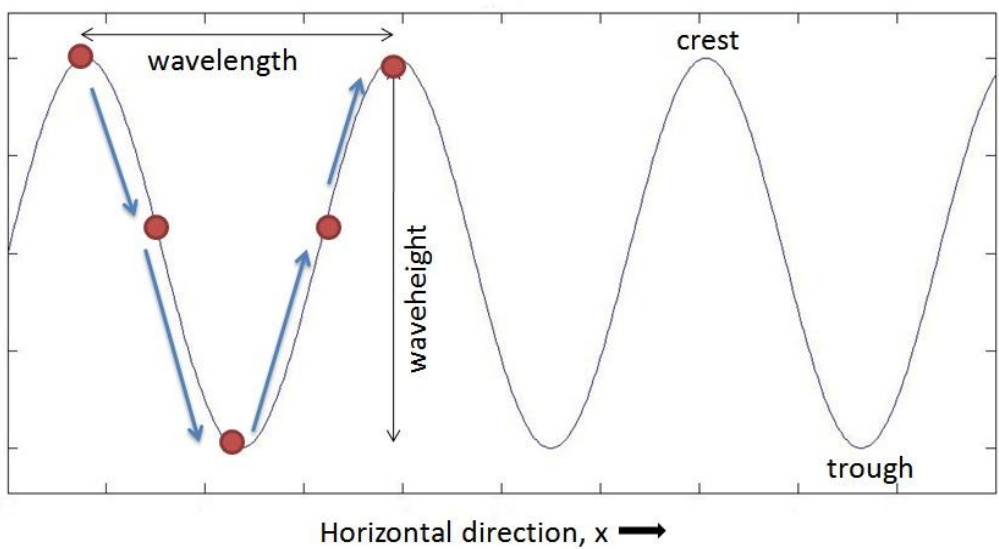

**Figure 1.2: Oscillations of water molecules propagate in horizontal direction.**

## 1.2 Governing equations of fluid mechanics

Before we proceed to the theory of water waves, we must at least have an understanding about the relevant governing equations in fluid mechanics. The governing equations serve the fundamental concept on the derivation of evolution equations. All of the equations are mostly related to each other but we will not discuss the derivation of the equations in details since this is merely a brief overview of water waves.

The fundamental concept of mass conservation is, in steady state, the rate of change of mass flowing into a control volume is equal to the rate of change of mass flowing out from the control volume. Therefore, the *mass conservation equation* (sometimes it is known as *continuity equation*) is given by (Euler, 1756)

$$\frac{\partial \rho}{\partial t} + \pmb{\nabla} \cdot (\rho \pmb{u}) = \frac{\partial \rho}{\partial t} + \frac{\partial (\rho u)}{\partial x} + \frac{\partial (\rho v)}{\partial y} + \frac{\partial (\rho w)}{\partial z} = 0, \qquad (1.1)$$

where $\rho$ is the density of the fluid and $\pmb{u} = (u, v, w)$ is the velocity of the fluid with velocity components $u, v$ and $w$.

For inviscid (an absence of viscous force) fluid, the motion of fluid is described by *Euler's equation* (Euler, 1756):

$$\frac{\partial \pmb{u}}{\partial t} + (\pmb{u} \cdot \pmb{\nabla})\pmb{u} = -\frac{1}{\rho}\pmb{\nabla}P + \pmb{F},$$

or in Cartesian form
$$\frac{\partial u}{\partial t} + \left(u\frac{\partial u}{\partial x} + v\frac{\partial u}{\partial y} + w\frac{\partial u}{\partial z}\right) = -\frac{1}{\rho}\frac{\partial P}{\partial x} + F_x,$$

$$\frac{\partial v}{\partial t} + \left(u\frac{\partial v}{\partial x} + v\frac{\partial v}{\partial y} + w\frac{\partial v}{\partial z}\right) = -\frac{1}{\rho}\frac{\partial P}{\partial y} + F_y, \qquad (1.2)$$

$$\frac{\partial w}{\partial t} + \left(u\frac{\partial w}{\partial x} + v\frac{\partial w}{\partial y} + w\frac{\partial w}{\partial z}\right) = -\frac{1}{\rho}\frac{\partial P}{\partial z} + F_z,$$

where $P$ is the pressure on the fluid and $\pmb{F} = (F_x, F_y, F_z)$ is the general body force per unit mass with components $F_x, F_y$ and $F_z$.

For mathematical simplicity, *Euler's equation* (1.1) is widely used in simple fluid mechanics problems. However, in reality, viscous force exists in every fluid substance. In order to model fluid flow closer to the realistic situation, a viscosity term is introduced into *Euler's equation* (1.2).

So we have *Navier-Stokes equation* (White, 1974):

$$\frac{\partial \boldsymbol{u}}{\partial t} + (\boldsymbol{u} \cdot \boldsymbol{\nabla})\boldsymbol{u} = -\frac{1}{\rho}\boldsymbol{\nabla}P + \boldsymbol{F} + \upsilon\boldsymbol{\nabla}^2\mathbf{u},$$

or
$$\frac{\partial u}{\partial t} + \left(u\frac{\partial u}{\partial x} + v\frac{\partial u}{\partial y} + w\frac{\partial u}{\partial z}\right) = -\frac{1}{\rho}\frac{\partial P}{\partial x} + F_x + \upsilon\left(\frac{\partial^2 u}{\partial x^2} + \frac{\partial^2 u}{\partial y^2} + \frac{\partial^2 u}{\partial z^2}\right),$$

$$\frac{\partial v}{\partial t} + \left(u\frac{\partial v}{\partial x} + v\frac{\partial v}{\partial y} + w\frac{\partial v}{\partial z}\right) = -\frac{1}{\rho}\frac{\partial P}{\partial y} + F_y + \upsilon\left(\frac{\partial^2 v}{\partial x^2} + \frac{\partial^2 v}{\partial y^2} + \frac{\partial^2 v}{\partial z^2}\right),$$

$$\frac{\partial w}{\partial t} + \left(u\frac{\partial w}{\partial x} + v\frac{\partial w}{\partial y} + w\frac{\partial w}{\partial z}\right) = -\frac{1}{\rho}\frac{\partial P}{\partial z} + F_z + \upsilon\left(\frac{\partial^2 w}{\partial x^2} + \frac{\partial^2 w}{\partial y^2} + \frac{\partial^2 w}{\partial z^2}\right),$$

(1.3)

where $\upsilon$ is the kinematic viscosity.

Now we can clearly see that with zero viscosity (where $\upsilon = 0$), the equation can be reduced to *Euler's equation* (1.2).

An important quantity of fluid flow is the *curl* of the velocity field, which is known as *vorticity* and denoted by $\boldsymbol{\omega} = (\omega_1, \omega_2, \omega_3)$. It describes the rotational motion of the fluid. For many flows, the vorticity is very small everywhere in the fluid. Therefore, for simplification purpose, we often make an assumption that $\boldsymbol{\omega} = \mathbf{0}$ which shows irrotationality. However, in the real life, real flows are rarely irrotational.

If $\boldsymbol{\omega}$ is introduced into *Euler's equation* (1.2), we obtain *Bernoulli's equation for unsteady flow* (Johnson, 1997), given as follows:

$$\frac{\partial \boldsymbol{u}}{\partial t} + \boldsymbol{\nabla}\left(\frac{1}{2}\boldsymbol{u} \cdot \boldsymbol{u} + \frac{P}{\rho} + \Omega\right) = \mathbf{u} \times \boldsymbol{\omega},$$

or
$$\frac{\partial u}{\partial t} + \frac{\partial}{\partial x}\left(\frac{1}{2}(u^2 + v^2 + w^2) + \frac{P}{\rho} + \Omega\right) = v\omega_3 - w\omega_2,$$

(1.4)

$$\frac{\partial v}{\partial t} + \frac{\partial}{\partial y}\left(\frac{1}{2}(u^2 + v^2 + w^2) + \frac{P}{\rho} + \Omega\right) = w\omega_1 - u\omega_3,$$

$$\frac{\partial w}{\partial t} + \frac{\partial}{\partial z}\left(\frac{1}{2}(u^2 + v^2 + w^2) + \frac{P}{\rho} + \Omega\right) = u\omega_2 - v\omega_1,$$

where $\Omega$ is a potential function so that $-\nabla\Omega = \boldsymbol{F}$.

There are two different cases for further examination. If the flow is steady, the velocity, pressure and potential function are independent of time. Therefore, it leads to the *Bernoulli's equation for steady flow* (Johnson, 1997), given as follows:

$$\frac{1}{2}\boldsymbol{u}\cdot\boldsymbol{u} + \frac{P}{\rho} + \Omega = \text{constant},$$

or

$$\frac{1}{2}(u^2 + v^2 + w^2) + \frac{P}{\rho} + \Omega = \text{constant}. \tag{1.5}$$

This *Bernoulli's equation* (1.5) describes the conservation of energy (kinetic energy + work done by pressure force + potential energy) for a steady and inviscid flow with vorticity.

For incompressible, irrotational but unsteady flow, we have $\boldsymbol{u} = \nabla\varphi$ for velocity potential, $\varphi(x, y, z, t)$. By substituting it into the *mass conservation equation* (1.1), we will get *Laplace's equation* (Johnson, 1997):

$$\nabla^2\varphi = \frac{\partial^2\varphi}{\partial x^2} + \frac{\partial^2\varphi}{\partial y^2} + \frac{\partial^2\varphi}{\partial z^2} = 0. \tag{1.6}$$

By quoting *Laplace's equation* (1.6) into *Euler's equation* (1.2), *Bernoulli's equation for unsteady flow* (1.4) (Johnson, 1997) thus becomes

$$\frac{\partial\varphi}{\partial t} + \frac{1}{2}\boldsymbol{u}\cdot\boldsymbol{u} + \frac{P}{\rho} + \Omega = f(t),$$

or

$$\frac{\partial\varphi}{\partial t} + \frac{1}{2}(u^2 + v^2 + w^2) + \frac{P}{\rho} + \Omega = f(t), \tag{1.7}$$

where $f(t)$ is an arbitrary function of integration.

## 1.3  Boundary conditions

There are various forms of boundary conditions that define the water-wave problem. We will discuss the theory of top and bottom surface boundary conditions in fluid. When the fluid surface is moving, it is always composed of fluid particles. Since the moving fluid surface does not involve any action of force, so it is called the *kinematic boundary condition*. The *kinematic free surface boundary condition* (*KFSBC*) (Johnson, 1997) is given by

$$\frac{\partial \eta}{\partial t} + \boldsymbol{\nabla}\varphi \cdot \boldsymbol{\nabla}\eta = \frac{\partial \eta}{\partial t} + \left(\frac{\partial \varphi}{\partial x}\frac{\partial \eta}{\partial x} + \frac{\partial \varphi}{\partial y}\frac{\partial \eta}{\partial y}\right) = \frac{\partial \varphi}{\partial z}, \qquad (1.8)$$

where $z = \eta(x, y, t)$ is the free surface elevation.

At the bottom of the fluid, if the fluid is treated as inviscid, the bottom topography becomes a surface of a fluid, therefore it is categorised as the *kinematic boundary conditions* as well (see Figure 1.3). It is often prescribed as *a priori* (Johnson, 1997).

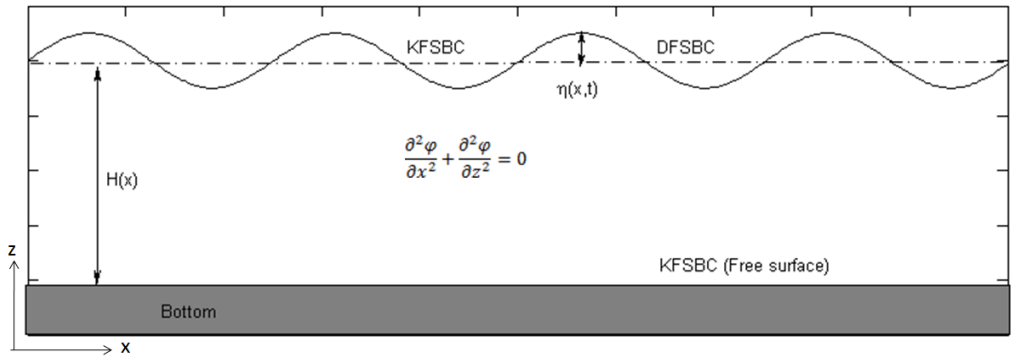

**Figure 1.3: Boundary conditions for inviscid water waves. KFSBC is due to the moving fluid particles and DFSBC is due to the atmospheric pressure.**

Revert back to water surface again. If the fluid is modelled as inviscid, the air in the atmosphere exerts only a pressure on the surface. The stress conditions in this case are called *dynamic boundary conditions*. If the viscosity is considered in the modelling of the fluid, tangential stresses from the surface tension will be introduced on top of the atmospheric pressure for the *dynamic free surface boundary conditions* (*DFSBC*). It is given by (Johnson, 1997)

$$\frac{\partial \varphi}{\partial t} + \frac{1}{2}(\nabla\varphi \cdot \nabla\varphi) + g\eta - \frac{T}{\rho}R = 0,$$

or

$$\frac{\partial \varphi}{\partial t} + \frac{1}{2}\left[\left(\frac{\partial \varphi}{\partial x}\right)^2 + \left(\frac{\partial \varphi}{\partial y}\right)^2 + \left(\frac{\partial \varphi}{\partial z}\right)^2\right] + g\eta - \frac{T}{\rho}R = 0,$$

(1.9)

where $g$ is the gravitational acceleration, $T$ is the surface tension and $R$ is the summation of the principal radii of curvature of the free surface.

For viscous flow, the boundary conditions on the water surface consist of KFSBC and DFSBC as well. But, at the bottom of the fluid, *bottom boundary conditions* (BBC, no slip conditions) (Johnson, 1997) where

$$\frac{\partial \varphi}{\partial z} = 0 \text{ at } z = 0,$$

must be imposed so that fluid particles in contact with the surface will move with the surface (see Figure 1.4).

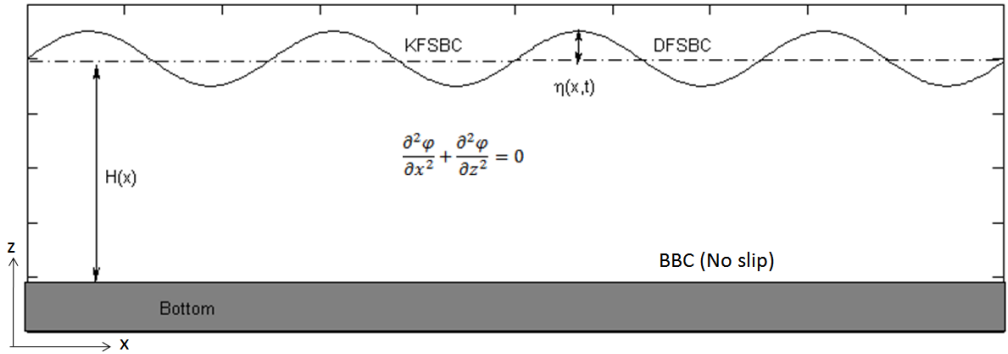

**Figure 1.4: Boundary conditions for viscous water waves.**

Thus, for a stationary bottom surface, the fluid velocity will be zero there. However, sometimes it could be modelled as a moveable bottom surface. A very good example is an earthquake modelling (Johnson, 1997).

## 1.4 Evolution equations

A wide range of nonlinear water wave equations in dispersive water waves will be discussed in this section. The motivation of discussing the equations is to provide better understandings on the different types of the nonlinear water wave equations. We assume that the water waves are of small amplitude and long wavelength, $\lambda$, compared to the water depth, $H$. More precisely, $\lambda > 7H$. (Rodríguez & Taboada-Vázquez, 2007).

We will begin with the shallow water waves. Shallow water waves are waves with amplitude dispersion and nonlinear. Consider an inviscid liquid of constant mean depth $H$ and constant density $\rho$. The free surface elevation above the undisturbed mean depth is $z = \eta(x, y, t)$, so that the free surface is at $z = H + \eta$ and $z = 0$ is the horizontal rigid bottom (see Figure 1.5).

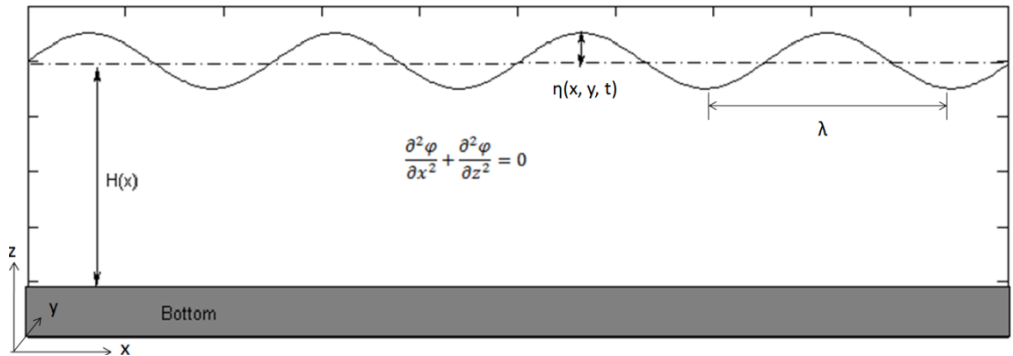

**Figure 1.5: Water waves on horizontal flat bottom.**

The *shallow water wave equations* are widely applicable in oceanic modelling especially near shores area. These are derived from the *Navier-Stokes equations* and *equations of conservation of mass* (Saint-Venant, 1843), in the case where the ratio of the wavelength to the water depth is very large $(\lambda > 7H)$ (Rodríguez & Taboada-Vázquez, 2007). The *shallow water wave equations* (two-dimensional) are given as follows (Boussinesq & Flament, 1886):

$$u_t + \varepsilon(uu_x + vv_z) + \eta_x - \frac{1}{2}\delta(u_{txx} + v_{txz}) = 0, \qquad (1.10)$$

$$v_t + \varepsilon(uv_x + vv_z) + \eta_z - \frac{1}{2}\delta(u_{txz} + v_{tzz}) = 0, \qquad (1.11)$$

$$\eta_t + [u(1+\varepsilon\eta)]_x + [v(1+\varepsilon\eta)]_z = \frac{\delta}{6}\left[\left(\frac{\partial^2 u}{\partial x^2}+\frac{\partial^2 u}{\partial z^2}\right)_x + \left(\frac{\partial^2 v}{\partial x^2}+\frac{\partial^2 v}{\partial z^2}\right)_z\right], \quad (1.12)$$

where *u* is the horizontal velocity (in the *x* direction), *v* is the vertical velocity (in the *z* direction), $\varepsilon$ is the dimensionless variable for surface wave amplitude to mean depth and $\delta$ is the dimensionless variable for mean depth to horizontal length.

Now consider the one-dimensional case and retaining both $\varepsilon$ and $\delta$ order terms so that these equations reduce to *Boussinesq's equations* (Boussinesq, 1872):

$$u_t + \varepsilon uu_x + \eta_x - \frac{1}{2}\delta u_{txx} = 0, \qquad (1.13)$$

$$\eta_t + [u(1+\varepsilon\eta)]_x - \frac{\delta}{6}u_{xxx} = 0. \qquad (1.14)$$

*Boussinesq's equations* describe the evolution of long water waves that travel in both positive and negative *x* directions and allow wave travelling simultaneously in the opposite direction (wave reflections).

Under small additional assumption that $\varepsilon$ and $\delta$ are of equal order and small, *Boussinesq's equations* can be reduced to *Korteweg-de Vries (KdV) equation* for wave that propagates in one horizontal dimension. The *KdV equation* (Korteweg & de Vries, 1895) is given as

$$\eta_t + 6\eta\eta_x + \eta_{xxx} = 0. \qquad (1.15)$$

It describes the motion of solitons and can be solved with exact solitons solution. Soliton is a self-reinforcing solitary wave that maintains its shape while travelling in constant speed (Zabusky & Kruskal, 1965). It is caused by the cancellation of nonlinear and dispersive effects in the medium where the dispersive effects are referred to as dispersion relations between frequency and the speed of the waves. The exact definitions of solitons (Drazin & Johnson, 1989) are

i. They are of permanent form.
ii. They are localised within a region.
iii. They can interact with each other and emerge from the collision unchanged, except for a phase shift.

Many exact solvable models have soliton solutions, including the *KdV equation* and the *Nonlinear Schrödinger (NLS) equation*. The soliton solutions (analytical solutions) are typically obtained by means of the inverse scattering transform (Gardner, Greene, Kruskal, & Miura, 1967).

The *KdV equation* was improved (Benjamin, Bona, & Mahony, 1972) for modelling long wave with small amplitude. The *Benjamin-Bona-Mahony (BBM) equation* (Benjamin, Bona, & Mahony, 1972) is given as follows:

$$\eta_t + \eta_x + \eta\eta_x - \eta_{xxt} = 0. \tag{1.16}$$

An extension of the *KdV equation* to motion in more than one-dimension was discovered by Kadomtsev & Petviashvili, (1970) who generalised the dispersion relation to give an extra term in the equation. The *Kadomtsev-Petviashvili (KP) equation (2-D KdV equation)* is given as

$$(\eta_t + 6\eta\eta_x + \eta_{xxx})_x + 3\lambda\eta_{yy} = 0, \tag{1.17}$$

where $\lambda = \pm 1$.

If a wave is slowly modulated as it propagates in a dispersive medium, most of the energy is confined in the neighbourhood. From the nonlinear dispersion relation involving both frequency and amplitude, the *NLS equation* is obtained as

$$i\psi_t + a\psi_{xx} + b|\psi|^2\psi = 0, \quad a, b \in \mathbb{R}, \tag{1.18}$$

where $\psi$ is the modulated wave packet. More detailed visualisation can be seen on the next chapter.

The *NLS equation* describes the evolution of the envelope of modulated wave group. Like the various form of the *KdV equation*, the *NLS equation* arises in many physical problems, including the nonlinear instability phenomena in optics, water waves and ocean waves. It was first derived by Zakharov (1968) via spectral method. A similar derivation of the NLS equation with an assumption of slowly varying bottom can be found in Djordjević & Redekopp (1978).

The extension of the *NLS equation* to soliton equation in 2+1 (space in *x*-direction and *y*-direction and time, *t*) dimension was introduced by Davey & Stewartson (1974) as the *Davey-Stewartson (DS) equations*:

$$i\psi_t + \psi_{xx} + \psi_{yy} = a|\psi|^2\psi + b\kappa\psi, \quad a, b \in \mathbb{R}, \tag{1.19}$$

$$\kappa_{xx} + c\kappa_{yy} = (|\psi|^2)_x, \quad c \in \mathbb{R}, \tag{1.20}$$

where $\kappa$ is the real field (mean-flow).

## 1.5   Outline of the thesis

In the following chapter, we will focus our discussion on the Nonlinear Schrödinger (NLS) equation since we are interested in solving it using a compact finite difference scheme. A more detailed introduction and the early development of the NLS equation will be presented. Then we will explain the physical interpretation of the NLS equation. The various forms of the NLS equation will be shown mathematically but we will concentrate mainly on the spatial NLS equation. Several examples on the evolution of the NLS equation will be presented for better visualisation.

In Chapter 3, we will discuss the development of numerical scheme to solve the NLS equation. Then, we will focus on the compact finite difference scheme. A scheme similar to Xie, Li, & Yi (2009) is adopted and modified to solve the spatial one-dimensional NLS equation with both constant and variable dispersive and nonlinear coefficients.

Then in Chapter 4, we will be looking at some numerical results produced using the compact finite difference scheme discussed in Chapter 3. Several topographies are considered and plotted in three-dimensional graphs to show the propagation of the envelope of the water wave signal along a horizontal spatial direction.

In the final chapter, we will conclude the thesis and give some recommendations for some possible future research directions.

# 2 Nonlinear Schrödinger Equation

## 2.1 Introduction

The nonlinear Schrödinger (NLS) equation is one of the most important universal nonlinear models in modern physics and applied mathematics. The NLS equation appears in quantum field theory (Arai, 1991), condensed matter (Bishop & Schneider, 1978), thermodynamic processes (Wheatly, Buchanan, Swift, Migliori, & Hofler, 1985), plasma physics (Stenflo & Yu, 1997), models of protein dynamics (Fordy, 1990), models of energy transfer in molecular systems (Bang, Christiansen, Rasmussen, & Gaididei, 1995), optics (Hasegawa & Tappert, 1973), (Kivshar & Konotop, 1989), (Bordon & Anderson, 1989), (Agrawal, 1995) and fluid mechanics (Benney & Newell, 1967), (Zakharov, 1968), (Hasimoto & Ono, 1972), (Ablowitz & Clarkson, 1991), (Mei & Li, 2004). In this thesis, the application of the NLS equation in modelling the evolution of water waves will be explained.

## 2.2 Physical interpretation of the NLS equation

In water waves, the NLS equation describes an evolution of the envelope of modulated wave groups, or sometimes are also known as wave packets (see Figure 2.1). The wave packets are localised group of waves which travel with the group velocity and occur when the waves are dispersive. The small axes on Figure 2.1 explain the wave patterns obtained based on different horizontal axes. Wave profile is recorded based on horizontal space axis whereas wave signal is acquired based on horizontal time axis.

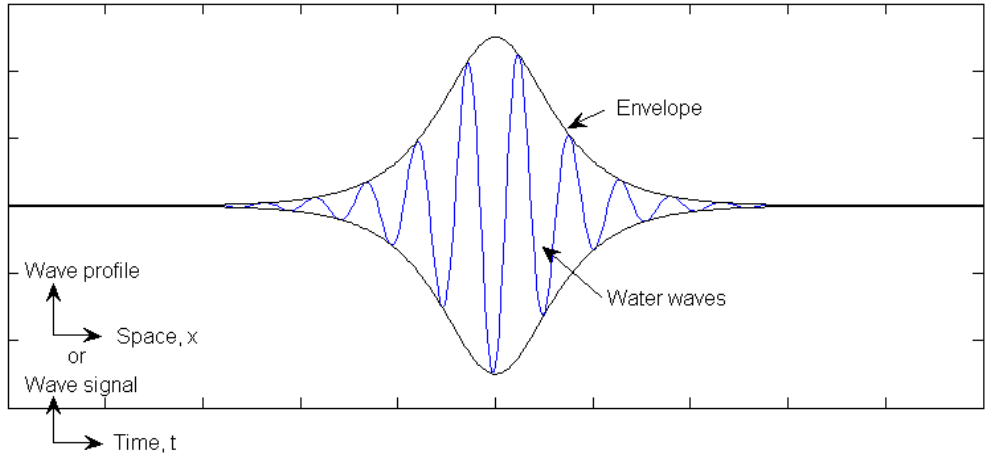

**Figure 2.1: Envelopes of modulated water waves group.**

The best known solution for NLS equation is of solitary waves form (Ruan & Chen, 2003). Solitary waves or often being called solitons are waves that remain in shapes and velocities upon interaction with other solitons by the cancellation of nonlinear and dispersive effects (Johnson, 1997).

There are several mathematical forms of the NLS equation. The most common NLS equation available in the literature is the temporal NLS equation. The temporal NLS equation illustrates the propagation of solitary waves in time scale. It is of imaginable model of equation, as similar to what our eyes see, as how we live, in a positive time domain.

In this thesis, we will focus mainly on the spatial NLS equation which describes the evolution of surface wave packets over a space domain because we are interested in the propagation of water wave signal over slowly varying topography (Benilov & Howlin, 2006). The physical appearance of the spatial NLS equation is beyond our imaginable model since the wave packets are recorded in signalling form, as opposed to what our eyes see, propagates in a space domain.

The NLS equation can be classified into NLS equation with constant coefficients and variable coefficients on the dispersive term and nonlinear term. For spatial NLS equation in water waves, the former describes the evolution of surface wave packets on a flat bottom while the latter describes the evolution of surface wave packets on varying topography. For optics, the spatial NLS equation with constant coefficients models the propagation of signal pulses whereas for the spatial NLS equation with non-constant coefficients, it records the amplification or absorption of signal pulses.

## 2.3 Mathematical model of the NLS equation

The NLS equation is considered as an initial value problem. We are interested in the one-dimensional NLS equation with a cubic nonlinearity particularly the spatial NLS equation with variable coefficients as the main objective is to understand the propagation of surface gravity waves over slowly varying topography.

### 2.3.1 Temporal cubic NLS equation

The general form of a one-dimensional temporal cubic NLS equation is given as follows:

$$i\psi_t + a\psi_{xx} + b\psi|\psi|^2 = 0, \qquad (2.1)$$

where *a* and *b* are arbitrary real constants and $\psi$ is the modulated wave packet. If *ab>0*, the NLS equation is of focusing type and has a bright soliton solution. Whereas if *ab<0*, the NLS equation is of defocusing type and has a dark soliton solution.

A simple numerical example of single soliton water waves from Xie, Li, & Yi (2009) where the NLS equation is given as:

$$i\frac{\partial \psi}{\partial t} + \frac{\partial^2 \psi}{\partial x^2} + 2|\psi|^2\psi = 0, \qquad (2.2)$$

with an initial condition of the form:

$$\psi(x, 0) = e^{2i(x-10)} \text{sech}(x - 10). \qquad (2.3)$$

The numerical results are shown in a three axes graph (Figure 2.2). We can see that a single soliton is travelling in an angle due to the exponential term in the initial condition.

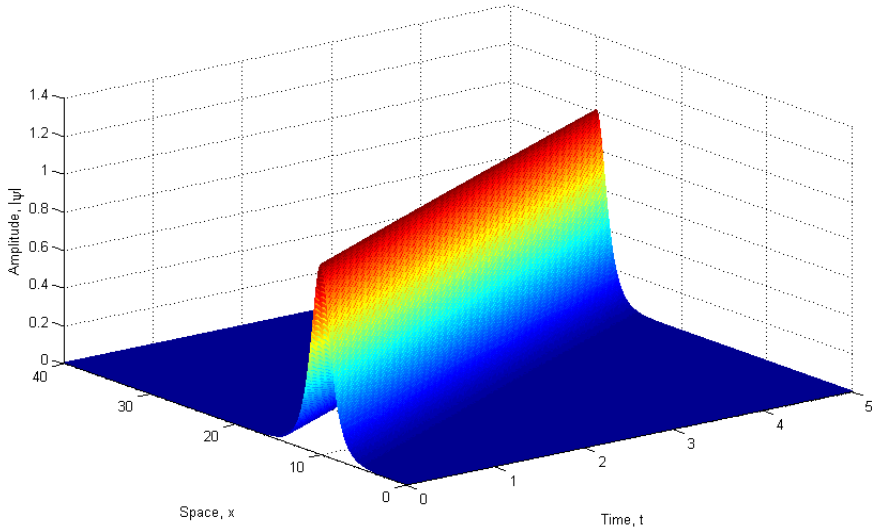

**Figure 2.2: A propagation of a single soliton with the temporal NLS equation (2.2) and an initial condition (2.3).**

Another numerical example of two solitons water wave from Xie, Li, & Yi (2009) with the NLS equation (2.2) but with a different initial condition given as

$$\psi(x, 0) = e^{0.5i(x-20)} \text{sech}(x - 20) + e^{0.05i(x-45)} \text{sech}(x - 45). \qquad (2.4)$$

The initial condition (2.4) yields a two-soliton solution, where the solitons are separated by a distance of 25 spatial units at $t = 0$. As shown in Figure 2.3, one of the solitons catches up with the other soliton and superposition occurred with a small phase shift resulting from the interaction. The solitons regain their shape upon collision despite a strongly nonlinear interaction, consistent with the soliton theory described in Section 1.4 and Section 2.2 on page 10 and 14 respectively.

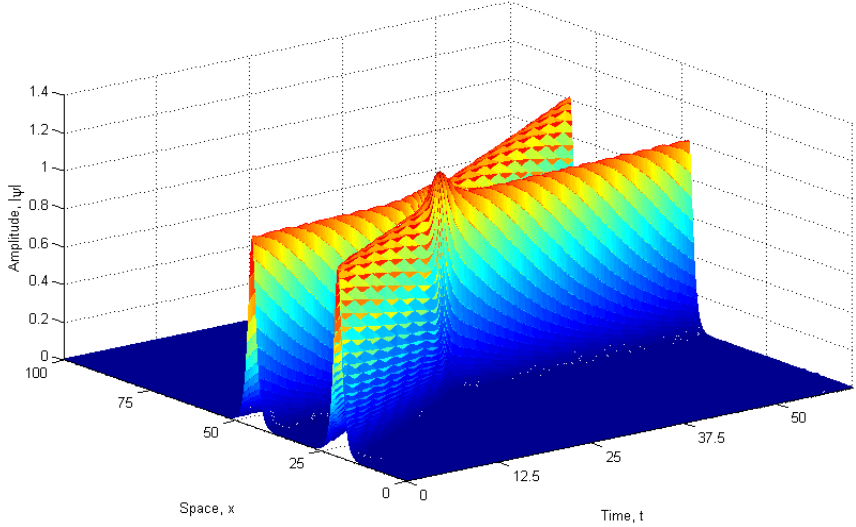

**Figure 2.3: An interaction of two solitons with the temporal NLS equation (2.2) and an initial condition (2.4).**

### 2.3.2 Spatial cubic NLS equation

For a one-dimensional spatial NLS equation with cubic nonlinearity, the equation with constant dispersive and nonlinear coefficients is now given as follows:

$$i\psi_x + \alpha\psi_{tt} + \beta\psi|\psi|^2 = 0, \qquad \alpha, \beta \in \mathbb{R}. \tag{2.5}$$

Notice that the role of time and space are now reversed.

We extend to a more complex one-dimensional spatial NLS equation with variable dispersive and nonlinear coefficients where it is given as follows:

$$i\psi_x + \alpha(x)\psi_{tt} + \beta(x)\psi|\psi|^2 = 0, \tag{2.6}$$

where *α(x)* and *β(x)* are slowly varying dispersion and nonlinear coefficients which depend on the position of *x*.

Now we consider an example from application of the NLS equation with non-constant coefficients in optics that can be found in the literature (Hao, Li, Li, Xue, & Zhou, 2004), where $\alpha(x) = \sin(x)/2$ and $\beta(x) = \sin x$, the NLS equation is given as:

$$i\frac{\partial \psi}{\partial x} + \frac{1}{2}\sin x \frac{\partial^2 \psi}{\partial t^2} + \sin x \, |\psi|^2 \psi = 0, \tag{2.7}$$

with a boundary condition of single soliton type:

$$\psi(0,t) = 1.5 e^{0.8i(t-10)} \operatorname{sech} 1.5(t-10). \tag{2.8}$$

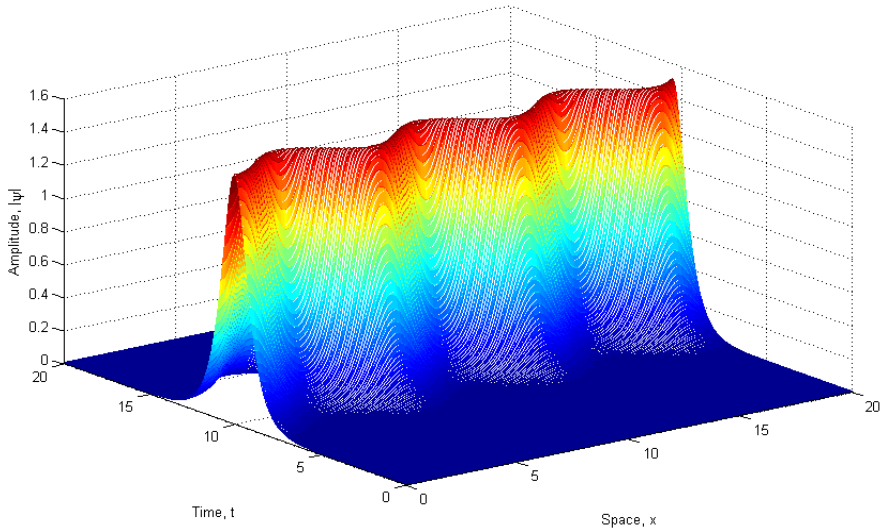

**Figure 2.4: An evolution plot of the spatial NLS equation with variable coefficients (2.7) and an initial condition (2.8).**

In this example, the plot shows a pulse of signal propagating in fibre optics with distributed dispersion and nonlinearity. In practical application, it is used to model the transmission of soliton signal. The soliton remains in shape despite moving in a sinusoidal direction, which is consistent with the soliton theory.

# 3 Numerical Scheme

## 3.1  Introduction

In this chapter, the development of numerical schemes to solve the nonlinear Schrödinger (NLS) equation will be presented. The compact finite difference scheme will be adopted from Xie, Li, & Yi (2009) to solve both one-dimensional NLS equation with constant and variable coefficients.

The initial-value problem of the one-dimensional NLS equation with constant dispersive and nonlinear coefficients can be solved analytically by using the inverse scattering transform method (Zakharov & Shabat, 1972), (Ablowitz & Segur, 1981), (Taha, 1991), (Taha, 1996).

There are abundant of investigations on numerical schemes to approximate the solution of the NLS equation based on either the finite difference (Delfour, Fortin, & Payre, 1981), (Taha & Ablowitz, 1984), (Zhang, 2005), the discontinuous Galerkin (Karakashian & Makridakis, 1998), (Xu & Shu, 2005), the spectral (Sulem, Sulem, & Patera, 1984), (Pathria & Morris, 1990) or the finite element (Tourigny & Morris, 1988), (Gardner, Gardner, Zaki, & Sahrawi, 1993) methods. Numerical results show that improper discretisation would bring on to numerical instability and non-convergent solutions.

According to Xie, Li, & Yi (2009), a compact finite difference scheme with high accuracy is proposed to solve numerically the temporal one-dimensional NLS equation with constant dispersive and nonlinear coefficients. A similar scheme is adopted and modified to solve the spatial one-dimensional NLS equation with both constant and variable dispersive and nonlinear coefficients.

## 3.2 The compact finite difference scheme

Here a brief introduction on the finite difference method will be discussed. Finite difference methods are numerical methods to approximate the solutions to differential equation using finite difference equations to approximate derivatives. The finite difference equation is derived from the expansion of Taylor's polynomial. The expansion of a function with an initial value of $a$ by Taylor's series is shown below:

$$f(a+h) = f(a) + \frac{f'(a)}{1!}h + \frac{f''(a)}{2!}h^2 + \cdots + \frac{f^{(n)}(a)}{n!}h^n + R_n(x), n \in \mathbb{N} \quad (3.1)$$

where $h$ denotes the small increment, $n!$ denotes the factorial of $n$, and $R_n(x)$ is a remainder term which has the form of $f^{(n+1)}(x)h^{n+1}/(n+1)!$ where $x$ is between $a$ and $a+h$. So, with some algebraic manipulation,

$$f'(a) = \frac{f(a+h) - f(a)}{h} + O(h). \quad (3.2)$$

When $h$ is small, approximation can be used as:

$$f'(a) \cong \frac{f(a+h) - f(a)}{h}. \quad (3.3)$$

Equation (3.3) is in fact a forward finite difference equation for the first derivative. Using this formula to replace derivative expressions in a differential equation, the solutions of the differential equation can be approximated.

### 3.2.1 Discretisation of the NLS equation with constant coefficients

The spatial one-dimensional NLS equation with constant coefficients is given as follows

$$i\psi_x + \alpha\psi_{tt} + \beta\psi|\psi|^2 = 0, \quad (3.4)$$

where $\alpha$ and $\beta$ are arbitrary real constants, $x$ is the horizontal space variable, $t$ is the positive time variable and $\psi$ is the wave packet amplitude.

We assume that the solution to equation (3.4) is bounded for a positive time period $[0,T]$ across a space interval of $[A,B]$. In order to model the physical condition that $\psi \to 0$ as $T \to \infty$, artificial initial and final conditions $\psi(x,0) = \psi(x,T) = 0$ are imposed. Under this assumption, equation (3.4) becomes a problem as follows:

$$i\psi_x + \alpha\psi_{tt} + \beta\psi|\psi|^2 = 0, \quad A \le x \le B, \quad 0 < t \le T,$$
$$\psi(0,t) = f(t), \quad 0 < t \le T, \quad (3.5)$$
$$\psi(x,0) = \psi(x,T) = 0, \quad A \le x \le B.$$

Let $h$ and $\phi$ be the discretisation parameters for the space and time mesh sizes respectively. The set of nodes $\{(x_j, t_n)\}$ is introduced into the discretised grid.

Let $\nu = \alpha\psi_{tt}$, equation (3.5) can be written as $\nu = -i\psi_x - \beta\psi|\psi|^2$. We denote $\nu_j^n = \nu(x_j, t_n)$, so

$$\nu_j^n = -i(\psi_x)_j^n - \beta\psi_j^n(|\psi|^2)_j^n. \quad (3.6)$$

Using a Taylor expansion with respect to space, we have

$$\psi_{j+1}^n = \psi_j^n + h(\psi_x)_j^n + \frac{h^2}{2}(\psi_{xx})_j^n + O(h^3), \quad (3.7)$$

$$\psi_{j-1}^n = \psi_j^n - h(\psi_x)_j^n + \frac{h^2}{2}(\psi_{xx})_j^n + O(h^3). \quad (3.8)$$

Subtracting (3.8) from (3.7) gives

$$\psi_{j+1}^n - \psi_{j-1}^n = 2h(\psi_x)_j^n + O(h^3),$$

$$(\psi_x)_j^n = \frac{\psi_{j+1}^n - \psi_{j-1}^n}{2h} + O(h^2). \quad (3.9)$$

For simplification purpose, we denote that

$$(\psi_j)_{\bar{x}}^n = \frac{\psi_{j+1}^n - \psi_{j-1}^n}{2h}. \quad (3.10)$$

So (3.10) becomes

$$(\psi_x)_j^n = (\psi_j)_{\hat{x}}^n + O(h^2). \tag{3.11}$$

Hence, substituting (3.11) into (3.6) yields

$$v_j^n = -i(\psi_j)_{\hat{x}}^n - \beta \psi_j^n (|\psi|^2)_j^n + O(h^2). \tag{3.12}$$

Now consider $v = \alpha \psi_{tt}$ and denote $v_j^n = v(x_j, t_n)$, so

$$v_j^n = \alpha (\psi_{tt})_j^n. \tag{3.13}$$

Using a Taylor expansion with respect to time, we have

$$\psi_j^{n+1} = \psi_j^n + \phi(\psi_t)_j^n + \frac{\phi^2}{2}(\psi_{tt})_j^n + \frac{\phi^3}{6}(\psi_{ttt})_j^n + \frac{\phi^4}{24}(\psi_{tttt})_j^n$$

$$+ \frac{\phi^5}{120}(\psi_{ttttt})_j^n + O(\phi^6), \tag{3.14}$$

$$\psi_j^{n-1} = \psi_j^n - \phi(\psi_t)_j^n + \frac{\phi^2}{2}(\psi_{tt})_j^n - \frac{\phi^3}{6}(\psi_{ttt})_j^n + \frac{\phi^4}{24}(\psi_{tttt})_j^n$$

$$- \frac{\phi^5}{120}(\psi_{ttttt})_j^n + O(\phi^6). \tag{3.15}$$

Adding (3.14) and (3.15) gives

$$\psi_j^{n+1} + \psi_j^{n-1} = 2\psi_j^n + \phi^2 (\psi_{tt})_j^n + \frac{\phi^4}{12}(\psi_{tttt})_j^n + O(\phi^6),$$

$$(\psi_{tt})_j^n = \frac{\psi_j^{n+1} - 2\psi_j^n + \psi_j^{n-1}}{\phi^2} - \frac{\phi^2}{12}(\psi_{tttt})_j^n + O(\phi^4). \tag{3.16}$$

For simplification purpose, we denote that

$$(\psi_j)_{\hat{t}\hat{t}}^n = \frac{\psi_j^{n+1} - 2\psi_j^n + \psi_j^{n-1}}{\phi^2}. \tag{3.17}$$

So, substituting (3.17) into (3.16) yields

$$(\psi_{tt})_j^n = (\psi_j)_{\hat{t}\hat{t}}^n - \frac{\phi^2}{12}(\psi_{tttt})_j^n + O(\phi^4). \tag{3.18}$$

Hence (3.13) now becomes

$$v_j^n = \alpha(\psi_j)_{\hat{t}\hat{t}}^n - \frac{\alpha\phi^2}{12}(\psi_{tttt})_j^n + O(\phi^4). \tag{3.19}$$

Perform second order partial derivative of $v = \alpha\psi_{tt}$ with respect to time, $t$ and let $(v_{tt})_j^n = v(x_j, t_n)$ give

$$(v_{tt})_j^n = \alpha(\psi_{tttt})_j^n. \tag{3.20}$$

Substituting (3.20) into (3.19) yields

$$v_j^n = \alpha(\psi_j)_{\hat{t}\hat{t}}^n - \frac{\phi^2}{12}(v_{tt})_j^n + O(\phi^4),$$

$$= \alpha(\psi_j)_{\hat{t}\hat{t}}^n - \frac{\phi^2}{12}\left[(v_j)_{\hat{t}\hat{t}}^n - \frac{\phi^2}{12}(v_{tttt})_j^n + O(\phi^4)\right] + O(\phi^4). \tag{3.21}$$

Since $\phi^4(v_{tttt})_j^n/144$ is of fourth order in $\phi$, so we neglect this term. (3.21) now becomes

$$v_j^n = \alpha(\psi_j)_{\hat{t}\hat{t}}^n - \frac{\phi^2}{12}(v_j)_{\hat{t}\hat{t}}^n + O(\phi^4). \tag{3.22}$$

Now equate (3.12) and (3.22)

$$-i(\psi_j)_{\hat{x}}^n - \beta\psi_j^n(|\psi|^2)_j^n + O(h^2) = \alpha(\psi_j)_{\hat{t}\hat{t}}^n - \frac{\phi^2}{12}(v_j)_{\hat{t}\hat{t}}^n + O(\phi^4). \tag{3.23}$$

Rearranging and substituting the relation of (3.17) into (3.23)

$$-i(\psi_j)_{\hat{x}}^n - \beta\psi_j^n(|\psi|^2)_j^n$$

$$= \alpha(\psi_j)_{\hat{t}\hat{t}}^n - \frac{1}{12}(v_j^{n+1} - 2v_j^n + v_j^{n-1}) + O(h^2 + \phi^4)$$

$$
\begin{aligned}
&= \alpha(\psi_j)_{\hat{t}\hat{t}}^n - \frac{1}{12}\{-i(\psi_j)_{\bar{x}}^{n+1} - \beta\psi_j^{n+1}(|\psi|^2)_j^{n+1} \\
&+ 2\left[i(\psi_j)_{\bar{x}}^n + \beta\psi_j^n(|\psi|^2)_j^n\right] - i(\psi_j)_{\bar{x}}^{n-1} - \beta\psi_j^{n-1}(|\psi|^2)_j^{n-1}\} + O(h^2 + \phi^4).
\end{aligned}
\tag{3.24}
$$

Rearranging the above expression gives

$$
\begin{aligned}
&\frac{i}{12}\left[(\psi_j)_{\bar{x}}^{n+1} + 10(\psi_j)_{\bar{x}}^n + (\psi_j)_{\bar{x}}^{n-1}\right] + \alpha(\psi_j)_{\hat{t}\hat{t}}^n \\
&+ \frac{\beta}{12}\left[\psi_j^{n+1}(|\psi|^2)_j^{n+1} + 10\psi_j^n(|\psi|^2)_j^n + \psi_j^{n-1}(|\psi|^2)_j^{n-1}\right] \\
&= O(h^2 + \phi^4).
\end{aligned}
\tag{3.25}
$$

Let $\Psi_j^n$ denotes the approximation of $\psi_j^n$, so (3.25) becomes

$$
\begin{aligned}
&\frac{i}{12}\left[(\Psi_j)_{\bar{x}}^{n+1} + 10(\Psi_j)_{\bar{x}}^n + (\Psi_j)_{\bar{x}}^{n-1}\right] + \alpha(\Psi_j)_{\hat{t}\hat{t}}^n \\
&+ \frac{\beta}{12}\left[\Psi_j^{n+1}(|\Psi|^2)_j^{n+1} + 10\Psi_j^n(|\Psi|^2)_j^n + \Psi_j^{n-1}(|\Psi|^2)_j^{n-1}\right] = 0.
\end{aligned}
\tag{3.26}
$$

Using central space averaging method where $\Psi_j$ is replaced by $(\Psi_{j+1} + \Psi_{j-1})/2$ in (3.26) gives

$$
\begin{aligned}
&\frac{i}{12}\left[(\Psi_j)_{\bar{x}}^{n+1} + 10(\Psi_j)_{\bar{x}}^n + (\Psi_j)_{\bar{x}}^{n-1}\right] + \frac{\alpha}{2}\left[(\Psi_{j+1})_{\hat{t}\hat{t}}^n + (\Psi_{j-1})_{\hat{t}\hat{t}}^n\right] \\
&+ \frac{\beta}{24}\{(\Psi_{j+1}^{n+1} + \Psi_{j-1}^{n+1})(|\Psi|^2)_j^{n+1} + 10(\Psi_{j+1}^n + \Psi_{j-1}^n)(|\Psi|^2)_j^n \\
&+ (\Psi_{j+1}^{n-1} + \Psi_{j-1}^{n-1})(|\Psi|^2)_j^{n-1}\} = 0.
\end{aligned}
\tag{3.27}
$$

From (3.10) and (3.17), we know that

$$
(\Psi_j)_{\bar{x}}^n = \frac{\Psi_{j+1}^n - \Psi_{j-1}^n}{2h},
\tag{3.28}
$$

$$
(\Psi_j)_{\hat{t}\hat{t}}^n = \frac{\Psi_j^{n+1} - 2\Psi_j^n + \Psi_j^{n-1}}{\phi^2}.
\tag{3.29}
$$

Thus, substituting (3.28) and (3.29) into (3.27) becomes

$$\frac{i}{24h}\{\Psi_{j+1}^{n+1} - \Psi_{j-1}^{n+1} + 10(\Psi_{j+1}^{n} - \Psi_{j-1}^{n}) + \Psi_{j+1}^{n-1} - \Psi_{j-1}^{n-1}\}$$

$$+\frac{\alpha}{2\phi^2}\{\Psi_{j+1}^{n+1} - 2\Psi_{j+1}^{n} + \Psi_{j+1}^{n-1} + \Psi_{j-1}^{n+1} - 2\Psi_{j-1}^{n} + \Psi_{j-1}^{n-1}\}$$

$$+\frac{\beta}{24}\{(\Psi_{j+1}^{n+1}+\Psi_{j-1}^{n+1})(|\Psi|^2)_j^{n+1} + 10(\Psi_{j+1}^{n}+\Psi_{j-1}^{n})(|\Psi|^2)_j^{n}$$

$$+(\Psi_{j+1}^{n-1}+\Psi_{j-1}^{n-1})(|\Psi|^2)_j^{n-1}\} = 0. \quad (3.30)$$

Rearranging gives

$$\frac{i}{24h}\{\Psi_{j+1}^{n+1} + 10\Psi_{j+1}^{n} + \Psi_{j+1}^{n-1} - \Psi_{j-1}^{n+1} - 10\Psi_{j-1}^{n} - \Psi_{j-1}^{n-1}\}$$

$$+\frac{\alpha}{2\phi^2}\{\Psi_{j+1}^{n+1} - 2\Psi_{j+1}^{n} + \Psi_{j+1}^{n-1} + \Psi_{j-1}^{n+1} - 2\Psi_{j-1}^{n} + \Psi_{j-1}^{n-1}\}$$

$$+\frac{\beta}{24}\{\Psi_{j+1}^{n+1}(|\Psi|^2)_j^{n+1} + 10\Psi_{j+1}^{n}(|\Psi|^2)_j^{n} + \Psi_{j+1}^{n-1}(|\Psi|^2)_j^{n-1}$$

$$+\Psi_{j-1}^{n+1}(|\Psi|^2)_j^{n+1} + 10\Psi_{j-1}^{n}(|\Psi|^2)_j^{n} + \Psi_{j-1}^{n-1}(|\Psi|^2)_j^{n-1}\} = 0. \quad (3.31)$$

Thus,

$$\frac{i}{24h}(\Psi_{j+1}^{n+1} + 10\Psi_{j+1}^{n} + \Psi_{j+1}^{n-1}) + \frac{\alpha}{2\phi^2}(\Psi_{j+1}^{n+1} - 2\Psi_{j+1}^{n} + \Psi_{j+1}^{n-1})$$

$$+\frac{\beta}{24}[\Psi_{j+1}^{n+1}(|\Psi|^2)_j^{n+1} + 10\Psi_{j+1}^{n}(|\Psi|^2)_j^{n} + \Psi_{j+1}^{n-1}(|\Psi|^2)_j^{n-1}]$$

$$= \frac{i}{24h}(\Psi_{j-1}^{n+1} + 10\Psi_{j-1}^{n} + \Psi_{j-1}^{n-1}) - \frac{\alpha}{2\phi^2}(\Psi_{j-1}^{n+1} - 2\Psi_{j-1}^{n} + \Psi_{j-1}^{n-1})$$

$$-\frac{\beta}{24}[\Psi_{j-1}^{n+1}(|\Psi|^2)_j^{n+1} + 10\Psi_{j-1}^{n}(|\Psi|^2)_j^{n} + \Psi_{j-1}^{n-1}(|\Psi|^2)_j^{n-1}]. \quad (3.32)$$

Imposing initial and final conditions (3.5) where $\psi(x,0) = \psi(x,T) = 0$, for a space interval of $x$, where $A \leq x \leq B$, (3.32) can be written as the following matrix equation for $j=1, 2, 3, ..., J$.

$$\left(\frac{i}{24h}[\mathbf{A}] + \frac{\alpha}{2\phi^2}[\mathbf{B}] + \frac{\beta}{24}[\mathbf{C}_j]\right)\Psi_{j+1} = \left(\frac{i}{24h}[\mathbf{A}] - \frac{\alpha}{2\phi^2}[\mathbf{B}] - \frac{\beta}{24}[\mathbf{C}_j]\right)\Psi_{j-1}, \quad (3.33)$$

where $[A]$, $[B]$ and $[C_j]$ are matrices of size $(N-1) \times (N-1)$,

$$[A] = \begin{bmatrix} 10 & 1 & 0 & \cdots & 0 \\ 1 & 10 & 1 & \ddots & \vdots \\ 0 & \ddots & \ddots & \ddots & 0 \\ \vdots & \ddots & 1 & 10 & 1 \\ 0 & \cdots & 0 & 1 & 10 \end{bmatrix}, \quad [B] = \begin{bmatrix} -2 & 1 & 0 & \cdots & 0 \\ 1 & -2 & 1 & \ddots & \vdots \\ 0 & \ddots & \ddots & \ddots & 0 \\ \vdots & \ddots & 1 & -2 & 1 \\ 0 & \cdots & 0 & 1 & -2 \end{bmatrix},$$

and

$$[C_j] = \begin{bmatrix} 10w_j^1 & w_j^2 & 0 & \cdots & 0 \\ w_j^1 & 10w_j^2 & w_j^3 & \ddots & \vdots \\ 0 & \ddots & \ddots & \ddots & 0 \\ \vdots & \ddots & w_j^{N-3} & 10w_j^{N-2} & w_j^{N-1} \\ 0 & \cdots & 0 & w_j^{N-2} & 10w_j^{N-1} \end{bmatrix},$$

where $w_j^n = |\Psi_j^n|^2$.

The compact finite difference scheme is similar to the original work of Xie, Li, & Yi (2009), except where the roles of time and space are reversed. Xie, Li, & Yi (2009) showed that the proposed scheme is of high accuracy (second and fourth order accurate in time and space) and preserves the conservation laws of charge and energy. However, it has limited application since the initial and final conditions have to be restricted to zero (refer equation 3.5) in order for this scheme to work.

### 3.2.2 Discretisation of the NLS equation with variable coefficients

The similar discretisation is performed and yields a similar matrix equation but with different dispersive and nonlinear coefficients.

For the spatial NLS equation with variable coefficients:

$$i\psi_x + \alpha(x)\psi_{tt} + \beta(x)\psi|\psi|^2 = 0. \tag{3.34}$$

The matrix equation for the spatial NLS equation with variable coefficient:

$$\left(\frac{i}{24h}[\mathbf{A}] + \frac{\alpha_j}{2\phi^2}[\mathbf{B}] + \frac{\beta_j}{24}[\mathbf{C}_j]\right)\Psi_{j+1} = \left(\frac{i}{24h}[\mathbf{A}] - \frac{\alpha_j}{2\phi^2}[\mathbf{B}] - \frac{\beta_j}{24}[\mathbf{C}_j]\right)\Psi_{j-1}. \quad (3.35)$$

So, the only difference from the discretised spatial NLS equation with constant coefficients (3.33) is the coefficients of matrices [**B**] and [**C**$_j$] where $\alpha = \alpha_j$ and $\beta = \beta_j$ which now depend on space variable, $x$.

## 3.3 The evolution equation for variable water depth

The NLS evolution equation for variable water depth is derived by Djordjević & Redekopp (1978) as

$$i[\psi_x + c_g^{-1}(x)\psi_t + \mu(x)\psi] + \alpha(x)\psi_{tt} + \beta(x)\psi|\psi|^2 = 0, \quad (3.36)$$

where $c_g$ is the group velocity.

The linear dispersion relation in (3.36) is given as

$$\omega^2 = k \tanh kH, \quad (3.37)$$

where $H = H(x)$ is the nondimensional water depth which depends on space, $x$.

The group velocity is given by

$$c_g(x) = \frac{d\omega}{dk}. \quad (3.38)$$

Therefore,

$$c_g(x) = \frac{1}{2\omega}(\tanh kH + kH\,\text{sech}^2\, kH). \quad (3.39)$$

and the dissipative coefficient

$$\mu(x) = \frac{1}{2c_g}\frac{dc_g}{dx}, \quad (3.40)$$

the dispersive coefficient

$$\alpha(x) = \frac{1}{2\omega c_g}\left(1 + \frac{2\omega H \tanh kH}{c_g} - \frac{H}{c_g^2}\right), \quad (3.41)$$

and the nonlinear coefficient

$$\beta(x) = \frac{1}{2\omega^3 c_g}\left(3k^4 + 2\omega^4 k^2 - \omega^8 - \frac{(2k\omega + k^2 c_g \operatorname{sech}^2 kH)^2}{H - c_g^2}\right). \quad (3.42)$$

In Djordjević & Redekopp (1978), they suggested that it is convenient to change (3.36) to a co-moving reference frame based on the conservation of the net energy flux by replacing $t$ with $\tau$ using the following relationship:

$$\tau = t - \int \frac{\mathrm{d}x}{c_g(x)}. \quad (3.43)$$

Then, in terms of $(x, \tau)$, NLS equation (3.36) becomes

$$i\left(c_g^{1/2}(x)\psi\right)_x + c_g^{1/2}(x)[\alpha(x)\psi_{\tau\tau} + \beta(x)\psi|\psi|^2] = 0. \quad (3.44)$$

Upon expanding

$$i\psi_x + i\frac{\left(c_g^{1/2}(x)\right)_x}{c_g^{1/2}(x)}\psi + \alpha(x)\psi_{\tau\tau} + \beta(x)\psi|\psi|^2 = 0, \quad (3.45)$$

(3.45) is similar to (3.34) but with an additional term of $i\left(c_g^{1/2}(x)\right)_x \psi / c_g^{1/2}(x)$. Therefore, the matrix equation (3.35) for numerical simulation needs to be modified.

So the initial boundary value problem now becomes

$$i\psi_x + i\frac{\left(c_g^{1/2}(x)\right)_x}{c_g^{1/2}(x)}\psi + \alpha(x)\psi_{\tau\tau} + \beta(x)\psi|\psi|^2 = 0,$$

$$A \leq x \leq B, \quad 0 < \tau \leq T, \quad (3.46)$$

$$\psi(0, \tau) = f(\tau), \quad 0 < \tau \leq T,$$

$$\psi(x, 0) = \psi(x, T) = 0, \quad A \leq x \leq B.$$

The initial boundary value problem of (3.46) has been examined asymptotically by Benilov & Howlin (2006) with the averaging method of dispersive and nonlinear coefficients. So now, we are interested in using a compact finite difference scheme to approximate (3.46) with variable dispersive and nonlinear coefficients.

The boundary condition is given by Benilov & Howlin (2006) as

$$\psi(0,\tau) = f(\tau) = \sqrt{\frac{2\alpha_0}{\beta_0}} \lambda \operatorname{sech}(\lambda\tau) e^{-\frac{i\nu\tau}{2\alpha_0}}, \qquad (3.47)$$

where $\alpha_0$ and $\beta_0$ are the value of $\alpha$ and $\beta$ when $x = 0$.

The reason that the numerical discretisation process is going to be discussed again is because there is an extra term in the NLS equation by Djordjević & Redekopp, (1978) as compared to the general NLS equation (3.5).

Let $\nu = \alpha(x)\psi_{\tau\tau}$, the NLS equation in equation (3.46) can be written as $\nu = -i\psi_x - i\left(c_g^{1/2}(x)\right)_x \psi(x)/c_g^{1/2} - \beta(x)\psi|\psi|^2$. We denote $\nu_j^n = \nu(x_j, \tau_n)$, so

$$\nu_j^n = -i(\psi_x)_j^n - i\left[\frac{\left(c_g^{1/2}\right)_x}{c_g^{1/2}}\right]_j \psi_j^n - \beta_j \psi_j^n (|\psi|^2)_j^n. \qquad (3.48)$$

Substituting (3.11) into (3.48) yields

$$\nu_j^n = -i(\psi_j)_{\hat{x}}^n - i\left[\frac{\left(c_g^{1/2}\right)_x}{c_g^{1/2}}\right]_j \psi_j^n - \beta_j \psi_j^n (|\psi|^2)_j^n + O(h^2). \qquad (3.49)$$

Equate (3.49) and (3.22)

$$-i(\psi_j)_{\hat{x}}^n - i\left[\frac{\left(c_g^{1/2}\right)_x}{c_g^{1/2}}\right]_j \psi_j^n - \beta_j \psi_j^n (|\psi|^2)_j^n + O(h^2)$$

$$= \alpha_j (\psi_j)_{\hat{\tau}\hat{\tau}}^n - \frac{\phi^2}{12}(\nu_j)_{\hat{\tau}\hat{\tau}}^n + O(\phi^4). \qquad (3.50)$$

Substituting the relation in (3.18) into (3.50) and rearranging gives

$$\frac{i}{12}\left[(\psi_j)_{\bar{x}}^{n+1} + 10(\psi_j)_{\bar{x}}^{n} + (\psi_j)_{\bar{x}}^{n-1}\right]$$

$$+\left[\frac{\left(c_g^{1/2}\right)_x}{c_g^{1/2}}\right]_j \psi_j^{n+1} + 10\left[\frac{\left(c_g^{1/2}\right)_x}{c_g^{1/2}}\right]_j \psi_j^{n} + \left[\frac{\left(c_g^{1/2}\right)_x}{c_g^{1/2}}\right]_j \psi_j^{n-1}$$

$$+\alpha_j (\psi_j)_{\bar{\tau}\tau}^{n}$$

$$+\frac{\beta_j}{12}\left[\psi_j^{n+1}(|\psi|^2)_j^{n+1} + 10\psi_j^{n}(|\psi|^2)_j^{n} + \psi_j^{n-1}(|\psi|^2)_j^{n-1}\right]$$

$$= O(h^2 + \phi^4).$$

(3.51)

Let $\Psi_j^n$ denotes the approximation of $\psi_j^n$ and using space averaging method

$$\frac{i}{24h}\{\Psi_{j+1}^{n+1} + 10\Psi_{j+1}^{n} + \Psi_{j+1}^{n-1}\}$$

$$+\frac{i}{24}\left[\frac{\left(c_g^{1/2}\right)_x}{c_g^{1/2}}\right]_j \{\Psi_{j+1}^{n+1} + 10\Psi_{j+1}^{n} + \Psi_{j+1}^{n-1}\}$$

$$+\frac{\alpha_j}{2\phi^2}\{\Psi_{j+1}^{n+1} - 2\Psi_{j+1}^{n} + \Psi_{j+1}^{n-1}\}$$

$$+\frac{\beta_j}{24}\{\Psi_{j+1}^{n+1}(|\Psi|^2)_j^{n+1} + 10\Psi_{j+1}^{n}(|\Psi|^2)_j^{n} + \Psi_{j+1}^{n-1}(|\Psi|^2)_j^{n-1}\}$$

$$= \frac{i}{24h}\{\Psi_{j-1}^{n+1} + 10\Psi_{j-1}^{n} + \Psi_{j-1}^{n-1}\}$$

$$-\frac{i}{24}\left[\frac{\left(c_g^{1/2}\right)_x}{c_g^{1/2}}\right]_j \{\Psi_{j-1}^{n+1} + 10\Psi_{j-1}^{n} + \Psi_{j-1}^{n-1}\}$$

$$-\frac{\alpha_j}{2\phi^2}\{\Psi_{j-1}^{n+1} - 2\Psi_{j-1}^{n} + \Psi_{j-1}^{n-1}\}$$

$$-\frac{\beta_j}{24}\{\Psi_{j-1}^{n+1}(|\Psi|^2)_j^{n+1} + 10\Psi_{j-1}^{n}(|\Psi|^2)_j^{n} + \Psi_{j-1}^{n-1}(|\Psi|^2)_j^{n-1}\}.$$

(3.52)

Imposing initial and final conditions from (3.46) where $\psi(x, 0) = \psi(x, T) = 0$, $A \leq x \leq B$, (3.52) can be written as the following matrix equation for $j = 1, 2, 3, ..., J$.

$$\left( \frac{i}{24h} [\mathbf{A}] + \frac{i}{24} \left[ \frac{\left(c_g^{1/2}\right)_x}{c_g^{1/2}} \right]_j [\mathbf{A}] + \frac{\alpha_j}{2\phi^2} [\mathbf{B}] + \frac{\beta_j}{24} [\mathbf{C}_j] \right) \Psi_{j+1}$$

$$= \left( \frac{i}{24h} [\mathbf{A}] - \frac{i}{24} \left[ \frac{\left(c_g^{1/2}\right)_x}{c_g^{1/2}} \right]_j [\mathbf{A}] - \frac{\alpha_j}{2\phi^2} [\mathbf{B}] - \frac{\beta_j}{24} [\mathbf{C}_j] \right) \Psi_{j-1}, \quad (3.53)$$

where $[\mathbf{A}]$, $[\mathbf{B}]$ and $[\mathbf{C}_j]$ are matrices of size $(N-1) \times (N-1)$ in (3.33).

The matrix equation (3.53) is dependent on space node ($j$). It can be solved by using iteration method to approximate the next node from each successive node. A boundary node, $\Psi_0$ and first node, $\Psi_1$ are needed to start the iteration loop.

Set $j = 0$ and let $\Psi_{-1} = \Psi_0$, we have

$$\left( \frac{i}{24h} [\mathbf{A}] + \frac{i}{24} \left[ \frac{\left(c_g^{1/2}\right)_x}{c_g^{1/2}} \right]_0 [\mathbf{A}] + \frac{\alpha_0}{2\phi^2} [\mathbf{B}] + \frac{\beta_0}{24} [\mathbf{C}_j] \right) \Psi_1$$

$$= \left( \frac{i}{24h} [\mathbf{A}] - \frac{i}{24} \left[ \frac{\left(c_g^{1/2}\right)_x}{c_g^{1/2}} \right]_0 [\mathbf{A}] - \frac{\alpha_0}{2\phi^2} [\mathbf{B}] - \frac{\beta_0}{24} [\mathbf{C}_j] \right) \Psi_0. \quad (3.54)$$

The entire compact finite difference scheme is implemented into MATLAB® to perform space marching loop calculations and the values of each calculation are stored in a matrix array. See Appendix (or the accompanying CD) for the MATLAB® code. So we will be using these codes to generate numerical results for different topography. Graphs will be shown in the next chapter to get a visualisation on the evolution of water waves signal.

# 4 Numerical Results

In this chapter, several topography shapes will be considered and the evolution of the wave packets will be approximated using the compact finite difference method in Chapter 3. The results obtained using compact finite difference scheme will be compared with the results published by Benilov, Flanagan, & Howlin (2005) and Benilov & Howlin (2006).

## 4.1 Comparison of the compact finite difference equation

In section 3.3, a compact finite difference scheme to solve the NLS equation (3.46) for water waves was discussed. Now, a comparison of the results produced using compact finite difference equation will be made with the results published by Djordjević & Redekopp (1978).

According to Benilov & Howlin (2006), the nondimensional depth of channel is chosen to be sinusoidal, where

$$H(x) = 1 + 0.6 \sin\left(\frac{x}{5}\right), \tag{4.1}$$

with an initial value of wavenumber,

$$k = 2 \text{ when } x = 0. \tag{4.2}$$

The simulation is carried out with the following parameters of the wave packet:

$$\lambda = 0.1 \text{ and } \nu = 0. \tag{4.3}$$

Then the relative amplitude of the wave packet is plotted across the space, *x*, with a relation of

$$\text{Relative amplitude of the packet } = \frac{\max\limits_{0<\tau<T}\{|\psi(x,\tau)|\}}{\max\limits_{0<\tau<T}\{\psi(0,\tau)\}}, \qquad (4.4)$$

to exaggerate the change in amplitude due the uneven but *smooth* (the horizontal scale of topography is much larger than the wavelength) topography.

So, by using the parameters (4.1), (4.2) and (4.3), the numerical simulation is carried out with step sizes for time interval, $\phi = 1$ with total time of 500 units and space interval, $h = 0.5$ with 0 and 200 as the lower and upper boundaries. The parameters are chosen with moderate computing time. Then the results are presented in Figure 4.1. The shape and the magnitude of the amplitude of the wave packet are identical to the results published by Benilov & Howlin (2006). The characteristics of the wave packet are identical where the packet amplitude does *two* oscillations per *one* oscillation of topography and the packet decays with decreasing depth and vice versa. This is due to the nonmonotonic dependence of the group velocity (and hence, the wave amplitude) on the channel's depth (Benilov & Howlin, 2006).

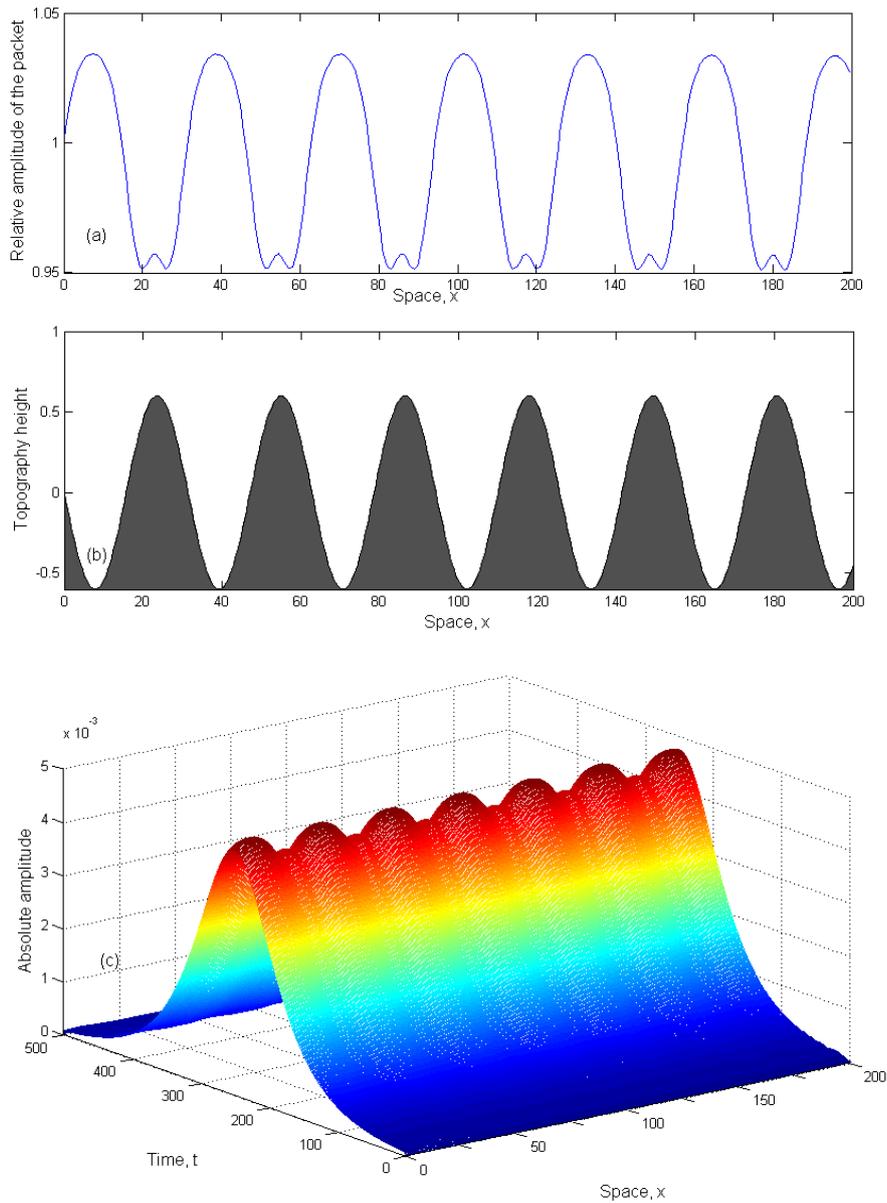

**Figure 4.1(a): A relative amplitude of the oscillating wave packet based on the initial amplitude across the spatial domain, x. (b) The depth of the channel (topography) across the spatial domain, x. (c) An evolution of wave signal across the spatial domain, x.**

The results produced by the compact finite difference method is comparable to the method used by Benilov & Howlin (2006). The next will be with the case considered by Benilov, Flanagan, & Howlin (2005) where the nondimensional water depth is given as

$$H(x) = 1 - 0.2 \tanh 0.002(x - 2000). \tag{4.5}$$

The rest of the parameters are identical to the previous simulation given by Benilov & Howlin (2006). The numerical simulation is carried out with a larger horizontal space scale from 0 to 6000 as lower and upper boundaries due to the nature of the equation (4.5).

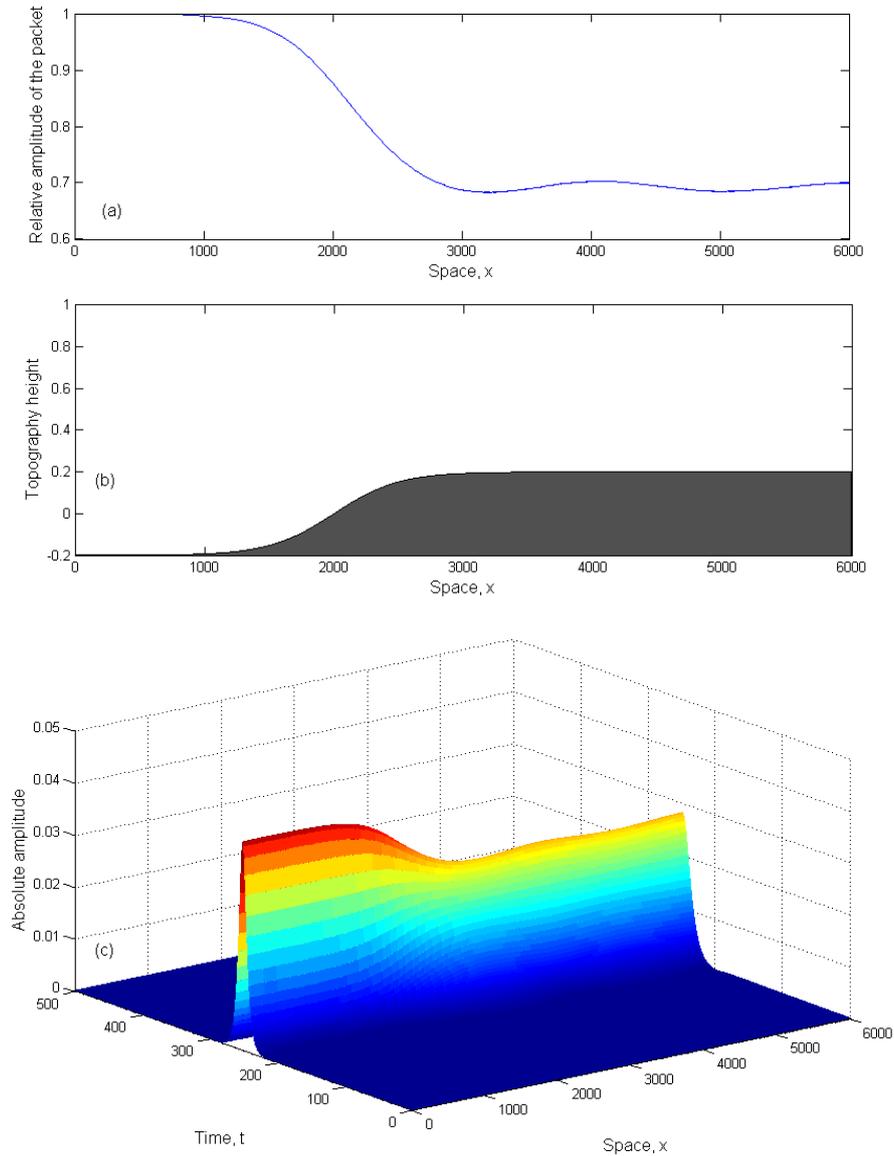

**Figure 4.2(a): A relative amplitude of the wave packet based on the initial amplitude across a gradually decreasing depth on the spatial domain, x. (b) The depth of the channel (topography) across the spatial domain, x. (c) The evolution of wave signal across the spatial domain, x.**

The shape and magnitude of the amplitude of the wave packet are identical to the results published by Benilov, Flanagan, & Howlin (2005). The above results model a packet of oceanic swell propagating over a continental shelf.

## 4.2 Examples of different topography

From the comparison between figures obtained from previous works and our computer simulation, the shape and magnitude of the amplitude of the wave packets are found to be almost identical. Now a few examples of different topography will be shown in order to understand the behaviour of the wave packets when they are travelling across a horizontal length scale.

In the following example, all the parameters remain unchanged except for the nondimensional water depth equation. We will start with a simple example of a bump on the flat bed of ocean topography. The nondimensional water depth is chosen as

$$H(x) = 1 - 0.3 \operatorname{sech}\left(\frac{x}{20} - 5\right). \tag{4.6}$$

The numerical experiment is carried out with the parameters of (3.47), (4.2), (4.3) and (4.6). The results are plotted with time interval, $\phi = 1$ for a period of 500 seconds and space interval, $h = 0.5$ for a space domain of 200 units as shown in Figure 4.3.

As we can see from Figure 4.3, the small bump on the flat bed of topography does affect a fairly small fall for the amplitude of the soliton travel across the horizontal spatial domain.

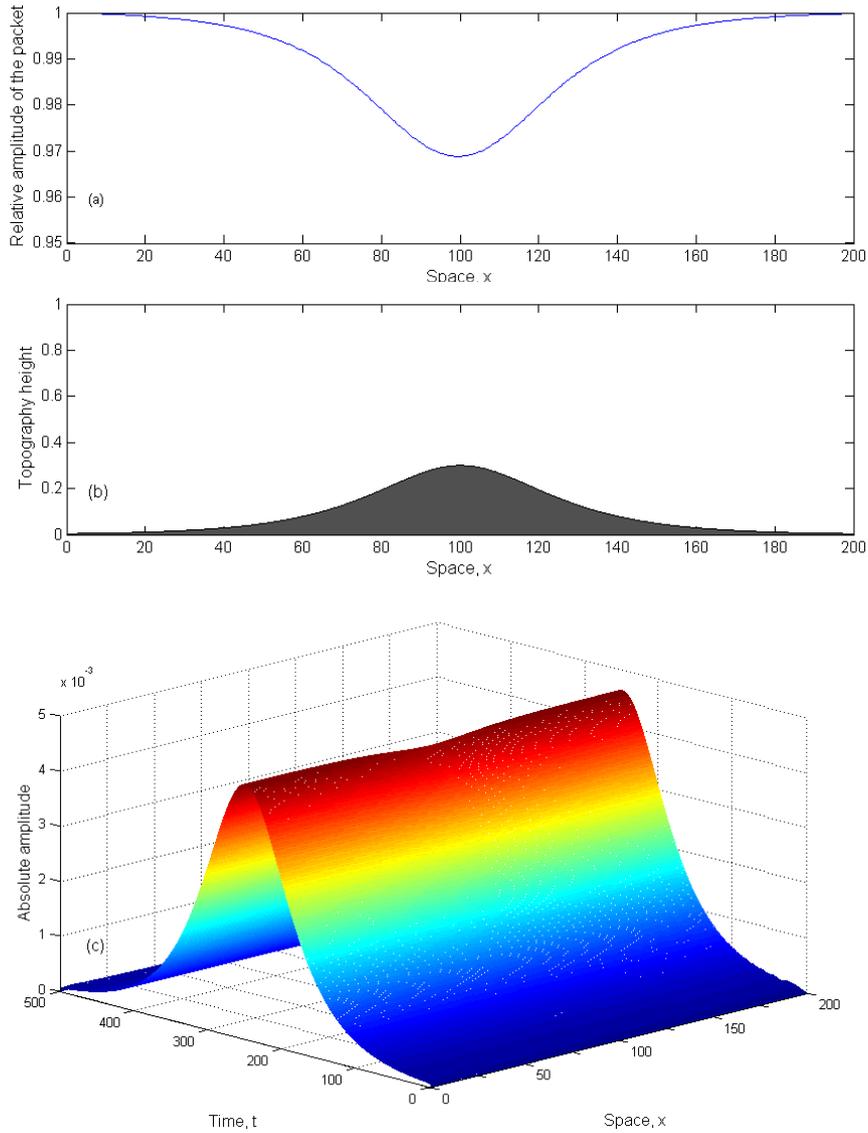

**Figure 4.3(a): A relative amplitude of the wave packet based on the initial amplitude across a small bump along the spatial domain, x. (b) The depth of the channel (topography) across the spatial domain, x. (c) An evolution of wave signal across the spatial domain, x.**

Next, we consider the opposite of (4.6). Instead of a bump, we consider a small dip on the flat bed of ocean. The nondimensional depth equation is chosen as

$$H(x) = 1 + 0.3 \operatorname{sech}\left(\frac{x}{20} - 5\right). \qquad (4.7)$$

The numerical simulation is carried out with the parameters of (3.47), (4.2), (4.3) and (4.7).

As we can see from Figure 4.4, the small dip on the flat bed of topography does affect a fairly small increase of the wave packet amplitude across the horizontal spatial domain.

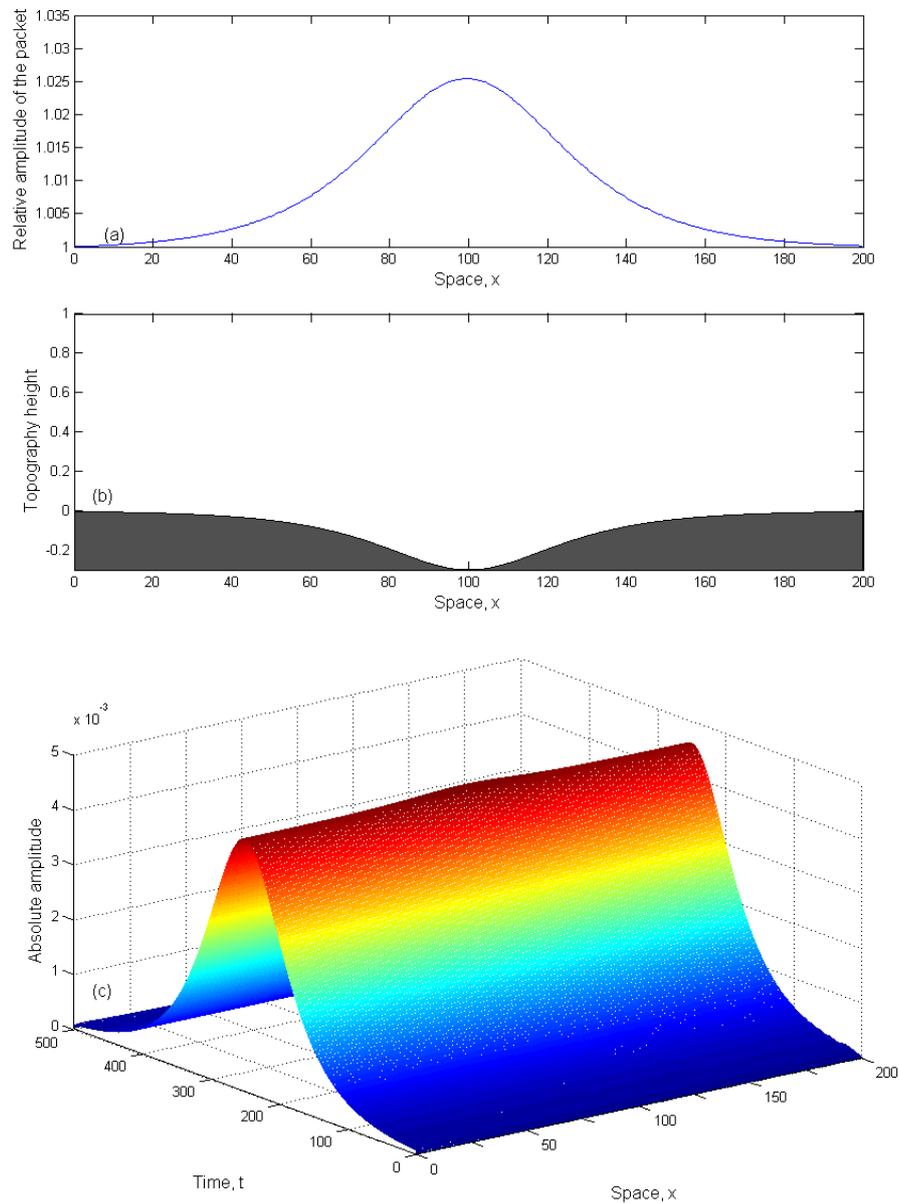

**Figure 4.4(a): A relative amplitude of the wave packet based on the initial amplitude across a small dip along the spatial domain, x. (b) The depth of the channel (topography) across the spatial domain, x. (c) An evolution of wave signal across the spatial domain, x.**

Then, we consider an example of gradual decrease but with substantial difference in topography height. The nondimensional depth equation is chosen as

$$H(x) = 1 + 0.2 \tanh 0.05(x - 100). \tag{4.8}$$

The numerical simulation is carried out with the parameters of (3.47), (4.2), (4.3) and (4.8).

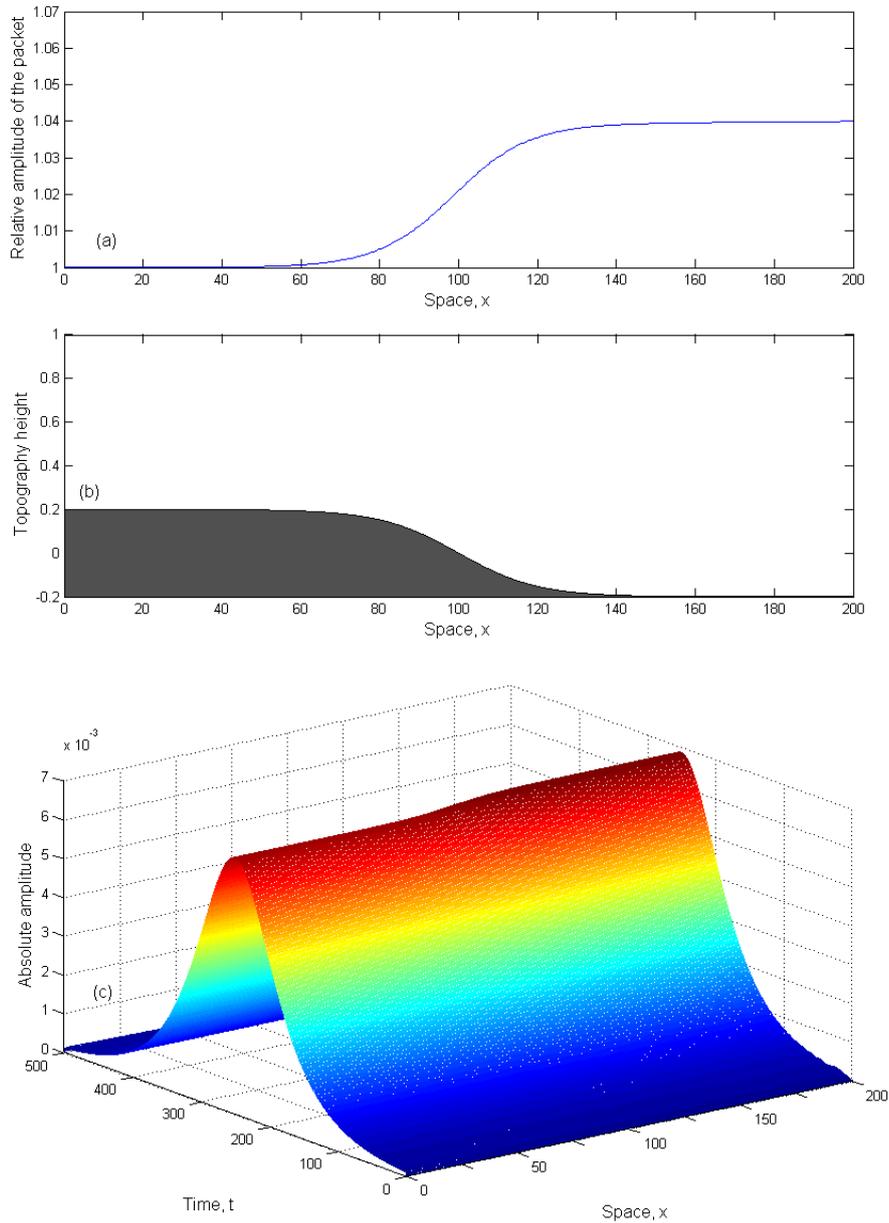

**Figure 4.5(a): A relative amplitude of the wave packet based on the initial amplitude across a gradual decrease topography height along the spatial domain, x. (b) The depth of the channel (topography) across the spatial domain, x. (c) An evolution of wave signal across the spatial domain, x.**

As shown in Figure 4.5, the wave packet amplitude increases with increase of the nondimensional depth. However the soliton shape remains identical to the initial soliton.

Now, we consider an oscillating decreasing nondimensional depth (increasing oscillating topography height) to model a more realistic shore area. The nondimensional depth equation is chosen as

$$H(x) = 1 - \left(0.1 \sin\left(\frac{x}{5}\right) + 0.0015x\right). \tag{4.9}$$

The numerical experiment is executed with the parameters (3.47), (4.2), (4.3) and (4.9).

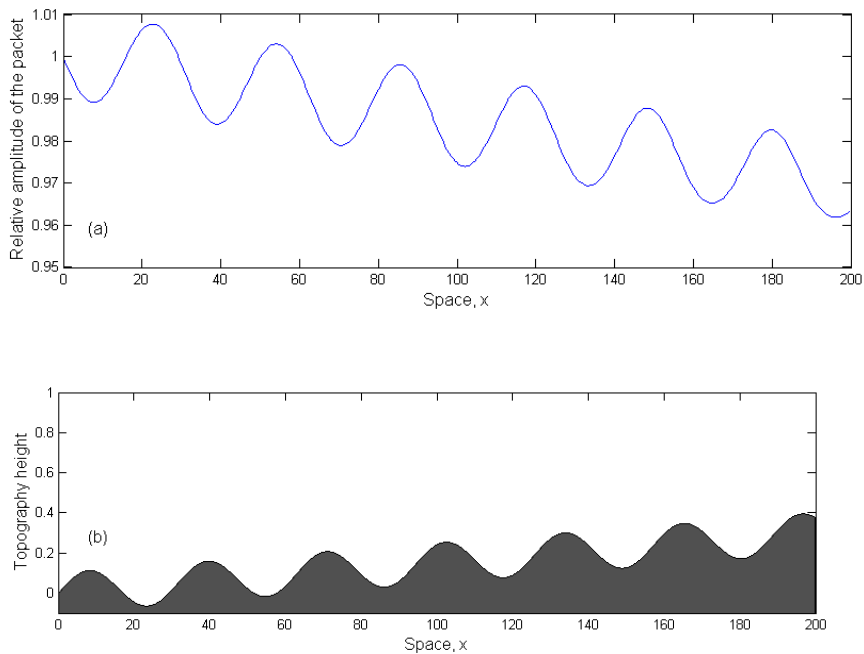

**Figure 4.6(a): A relative amplitude of the oscillating wave packet based on the initial amplitude across the slowly increasing oscillating topography height along the spatial domain, x. (b) The depth of the channel (topography) across spatial domain, x.**

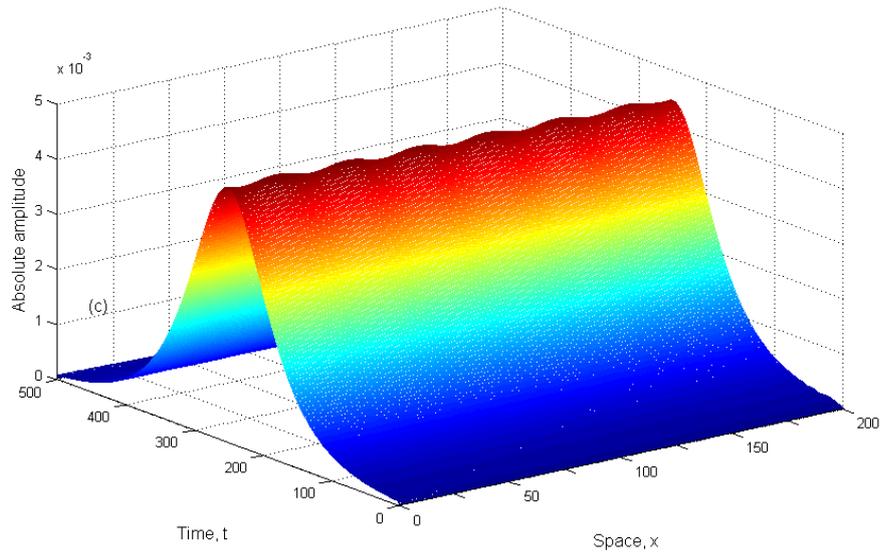

**Figure 4.7 :(c) An evolution of wave signal across spatial domain, x.**

The results show that with increasing topography height, the amplitude of the wave packet is decreasing (see Figure 4.6). However, the wave packet is still remained in shape, which is consistent with soliton theory that we have discussed in Sections 1.4 and 2.2.

The final example shows that sudden decrease of nondimensional depth. This models the oceanic cliff underwater. The nondimensional depth equation is chosen as

$$H(x) = 1 - \left[0.1 \cos\left(\frac{3x}{20}\right) + 0.3 \tanh(x - 100)\right]. \qquad (4.10)$$

The numerical experiment is executed with the parameters (3.47), (4.2), (4.3) and (4.10).

As we can see from Figure 4.8, the sudden increase of the topography height decreases the amplitude of the wave packet. The oscillating topography shows that the wave packet decays with decreasing depth and vice versa.

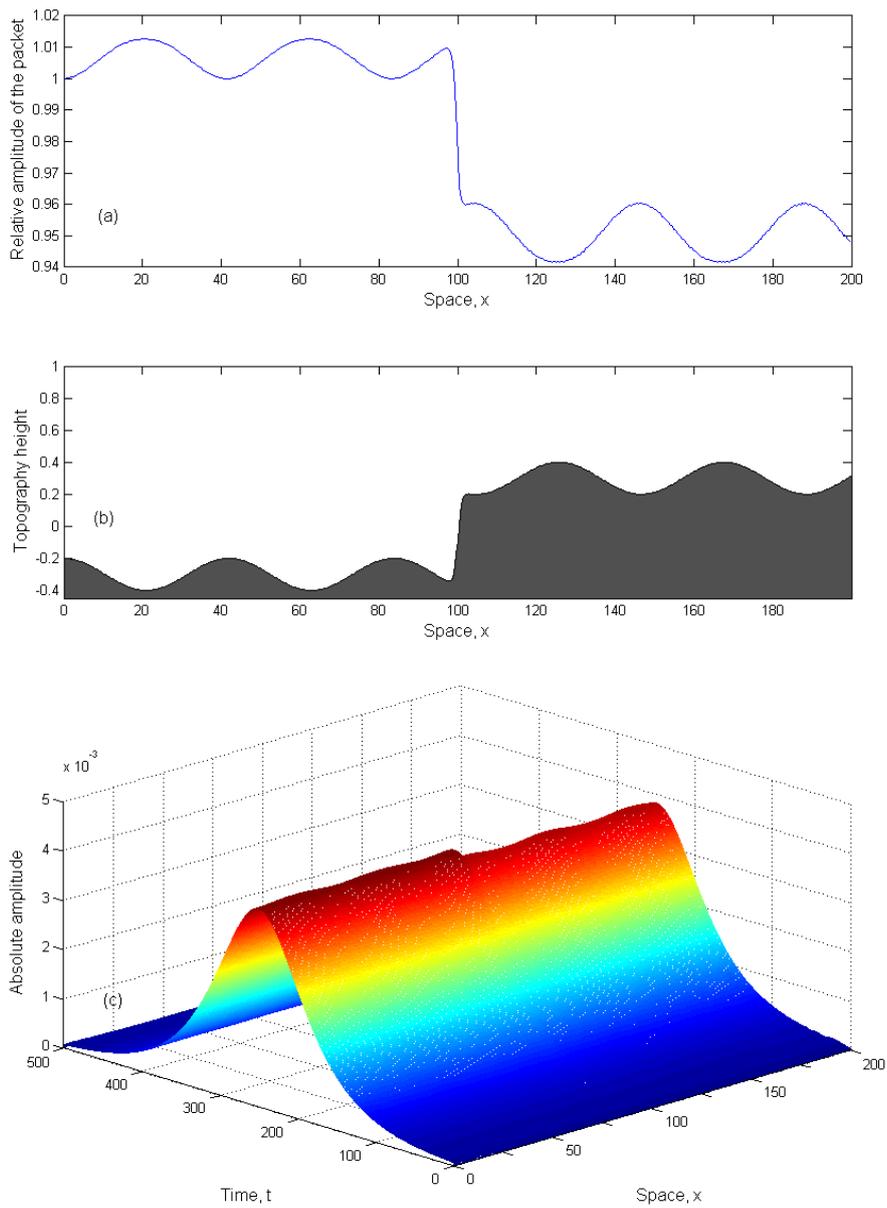

**Figure 4.8(a): A relative amplitude of the wave packet based on the initial amplitude across a sudden increasing and oscillating topography height along the spatial domain, x. (b) The depth of the channel (topography) across spatial domain, x. (c) An evolution of wave signal across spatial domain, x.**

# 5 Conclusion and Further Research

## 5.1   Conclusion

Most of the compact finite difference schemes available in the existing literature are used to solve the temporal nonlinear Schrödinger (NLS) equation. To the best of our knowledge, no work has used compact finite difference method on the spatial NLS equation with variable coefficients. In this work, the compact finite difference scheme is adopted from Xie, Li, & Yi (2009). It has been modified in order to solve the spatial NLS equation with variable dispersive and nonlinear coefficients. The results are compared with the results of Benilov, Flanagan, & Howlin (2005) and Benilov & Howlin (2006) which they used asymptotic analysis.

Several other examples have been shown to get a better understanding and visualisation on water wave packet signal travels along an uneven but *smooth* (the horizontal scale of topography is much larger than the wavelength) topography. It is observed that the wave envelope signal propagates opposed to topography height.

The possibility of shaping topography near shores area can be considered to generate desired wave signal for surfing purpose or to minimise the effect of high amplitude waves which bring disaster to the civilisation at the seaside (Goto, Takahashi, Oie, & Imamura, 2010).

## 5.2   Future Research

The higher order modified NLS equation, also known as Dysthe's equation (Dysthe, 1979) can be investigated numerically to model the wave propagation in higher accuracy. This equation has been implemented by Lo & Mei (1985) and has been improved to include broader spectral bandwidth by Trulsen & Dysthe (1996). Recently, Shermer & Dorfman (2008) have successfully obtained a good agreement between numerical simulation and experimental results. It will be interesting to solve this modified NLS equation using compact finite difference scheme. We can also consider impose different initial condition to verify the robustness of the numerical solution.

Finally, the obtained solution can be implemented to investigate the possibility of shaping topography near shores area to minimize the effect of extreme wave such as tsunami and thus mitigate wave hit problem.

# Appendix

We present the MATLAB® code to solve equation (3.47) using the compact finite difference equation (3.71) as follows:

----------------------------------------------------------------

```
clear,
clc; close all; %Clear the memory

tau   = 1;     %Time interval
t     = 500;   %Total time
lowx  = 0;     %Lower boundary
highx = 200;   %Upper boundary
h     = 0.5;   %Space interval
lambda= 0.1;   %Wavelength
v     = 0;     %Local curvature

%Create a tridiagonal matrix [A]
A = diag(10*ones((t/tau)-1,1),0) + diag(ones(((t/tau)-2),1),-1)
+ diag (ones(((t/tau)-2),1),1);
%Create a tridiagonal matrix [B]
B = diag(-2*ones((t/tau)-1,1),0) + diag(ones(((t/tau)-2),1),-1)
+ diag (ones(((t/tau)-2),1),1);

%Allocate time in [T]
T   = [0+tau:tau:t-tau]';
%Boundary value of \psi
U0 = sqrt(2*alpha(0)/beta(0))*lambda*sech(lambda*(T-
500)).*exp(1i*v*(T-500)/(2*(-alpha(0))));
%Squared absolute value of initial value of \psi
w0 = (abs(U0)).^2;
%Multiply the diagonal of matrix [A] with w0
C0 = A.*diag(w0);

%First node
U1 = (1i*sqrt(cg(h))*A/(24*h) + 1i*cgx(h)*A/24 +
sqrt(cg(h))*alpha(h)*B/(2*tau^2) + sqrt(cg(h))*beta(h)*C0/24)...
    \(1i*sqrt(cg(h))*A/(24*h) - 1i*cgx(h)*A/24 -
sqrt(cg(h))*alpha(h)*B/(2*tau^2) -
sqrt(cg(h))*beta(h)*C0/24)*U0;

%Spatial domain
maxcount = (highx-lowx) / h;

%Create an array for \psi
u = zeros((t/tau) -1, maxcount);

%Store the boundary value of \psi
u(:,1) = U0;
%Store the first node of \psi
u(:,2) = U1;

%Iteration loop
for iter = 1 : maxcount - 2
    w = (abs(u(:,iter+1))).^2;
    D = A*diag(w);
```

```matlab
    E = (1i*sqrt(cg((iter+1)*h))*A/(24*h) + 1i*cgx((iter+1)*h)*A/24 ...
        + sqrt(cg((iter+1)*h))*alpha((iter+1)*h)*B/(2*tau^2) + sqrt(cg((iter+1)*h))*beta((iter+1)*h)*D/24;
    F = (1i*sqrt(cg((iter+1)*h))*A/(24*h) - 1i*cgx((iter+1)*h)*A/24 ...
        - sqrt(cg((iter+1)*h))*alpha((iter+1)*h)*B/(2*tau^2) - sqrt(cg((iter+1)*h))*beta((iter+1)*h)*D/24;

    u(:,iter+2) = E\F*u(:,iter);
end;

%Set lower and upper boundary value to be zero
y = vertcat(zeros(1,maxcount),u,zeros(1,maxcount));

%Absolute value
Y = abs(y);

%Plot relative amplitude of the envelope
r=max(abs(u))/max(U0);
x=[lowx:h:highx-h];
plot(x,r);
figure

%Plot evolution of wave packet
mesh(Y);
rotate3D on;
%-------------------------------------------------------------------------

function [out] = alpha(x)
%Dispersive coefficient \alpha(x)
out = (1+ (2*omega(x).*H(x).*tanh(H(x)*k(x))./cg(x))-(H(x)./(cg(x).^2)))/(2*omega(x).*cg(x));

%-------------------------------------------------------------------------

function [out] = beta(x)
%Nonlinear coefficient \beta(x)
out = (3*k(x)^4+2*omega(x)^4*k(x)^2-omega(x)^8-((2*k(x)*omega(x)+k(x)^2*cg(x)*(sech(k(x)*H(x)))^2)^2)...
    /(H(x)-cg(x)^2))/(2*omega(x)^3*cg(x));
%-------------------------------------------------------------------------

function f = wavenumber(x)
%Calculate wavenumber, k based on newton-rhapson method
omega0=sqrt(2*tanh(2*H(0)));
k0 = 2;
f = 1;
err = 1e-6;
while abs(k0 - f) > err
    k0 = f;
    f = k0 - 2*(sqrt(k0*tanh(k0*H(x))) - omega0)*sqrt(k0*tanh(k0*H(x)))...
        /(tanh(k0*H(x)) + k0*H(x)*(1 - tanh(k0*H(x))^2));
end

%-------------------------------------------------------------------------
```

```matlab
function [out] = omega(x)
%\Omega value based on the linear dispersion relation at (3.38)
out=sqrt(k(x)* tanh(H(x)*k(x)));
```
-------------------------------------------------------------------------------------------------

```matlab
function [out] = k(x)
%Wavenumber, k
out=wavenumber(x);
```
-------------------------------------------------------------------------------------------------

```matlab
function [out] = cg(x)
%Group velocity \cg(x)
out=(tanh(k(x)*H(x))+k(x)*H(x).*(sech(k(x)*H(x)).^2))./(2*omega(x));
```
-------------------------------------------------------------------------------------------------

```matlab
function [out] = H(x)
%Delete where appropriate, only one function is allowed per iteration
%Nondimensional depth (4.12)
out=1+0.6*sin(x/5);

%Nondimensional depth (4.16)
out=1-0.2*tanh(0.002*(x-2000));

%Nondimensional depth (4.17)
out = 1-0.3*sech(x/20-5);

%Nondimensional depth (4.18)
out = 1+0.3*sech(x/20-5);

%Nondimensional depth (4.19)
out=1+0.2*tanh(0.05*(x-100));

%Nondimensional depth (4.20)
out=1-(0.1*sin(x/5)+0.0015*x);

%Nondimensional depth (4.21)
out=1-(0.1*cos(0.15*x)+0.3*tanh(x-100));
```
-------------------------------------------------------------------------------------------------

```matlab
function [out] = cgxf(x)
%Differentiate the cg function
%Enable symbolic function (syms x) before execution
%H
H=1+0.3*sech(x/20-5);
%**(H(0))
k = 2 - 2*(sqrt(2*tanh(2*H)) - sqrt(2*tanh(2*(1.004))))*sqrt(2*tanh(2*H))...
        /(tanh(2*H) + 2*H*(1 - tanh(2*H)^2));
omega=sqrt(k* tanh(H*k));
cg=(tanh(k*H)+k*H*(sech(k*H))^2)/(2*omega);
cgf=sqrt(cg);

out= diff(cgf,x);

%change H and **(value of H(0)) for different topography
```
-------------------------------------------------------------------------------------------------